\documentclass[12pt]{article}

\newcommand{\x}{arXiv:}

\usepackage{graphicx}

\begin{document}
\thispagestyle{empty}
\begin{center}

\null \vskip-1truecm \vskip2truecm

{\Large{\bf \textsf{The Arrow Of Time In The Landscape} }}

\vskip1truecm

{\large \textsf{Brett McInnes}}

\vskip1truecm

\textsf{\\  National
  University of Singapore}

\textsf{email: matmcinn@nus.edu.sg}\\

\end{center}
\vskip1truecm \centerline{\textsf{ABSTRACT}} \baselineskip=15pt
\medskip
The future is unlike the past. This is an absolutely basic
observation, as basic as the observation that ``things fall". We
expect that a real understanding of the Arrow of Time
---$\,$ the temporal ``direction" defined by the physical
differences between the future and the past ---$\,$ will arise
only in the context of some deep theory, such as string theory.
Conversely, we argue here that such an understanding is urgently
required if recent string-theoretic ideas about cosmology are to
be made to function.

In this work, we explain in detail why it is so difficult to find
a satisfactory theory of the Arrow of Time. There are several
difficulties, but the key problems arise from Roger Penrose's
observation that the Arrow is ultimately connected with the
extremely ``non-generic" character of the spatial geometry of the
earliest Universe. We argue that the explanations of the Arrow
proposed hitherto are unsatisfactory, because they do not address
this basic point. In any case, none of them applies to the string
Landscape, in which the nucleation of baby universes is postulated
to ``populate" the multiverse. Here we argue that baby universes
can only have an Arrow if they \emph{inherit} one; the problem of
explaining the Arrow is thus reduced to explaining it in the case
of the original universe. Motivated by a recent formulation of
``creation from nothing" in the context of string theory, we
propose that the original universe was created along a spacelike
surface with the topology of a torus. Using deep results in global
differential geometry, we are able to show that the geometry of
this surface had to be non-generic. This geometric ``specialness"
is communicated to matter through the inflaton, the basic physical
field postulated by the theory of Inflation. Thus we have a theory
of the Arrow which is intrinsically geometric, which incorporates
Inflation, and which allows universes in the Landscape to begin
with physically acceptable initial conditions.

\newpage

\addtocounter{section}{1}
\section* {\large{\textsf{1. 10$^{500}$ ---$\,$ You Call That ``Large"}?}}
As is well known ---$\,$ some would say ``notorious" ---$\,$
string theory admits a ``very large" number of internally
consistent basic models of the Universe: this is the string
Landscape \cite{kn:schellekens}\cite{kn:landscape}\cite{kn:TASI}.
The number usually cited, simply for definiteness, is 10$^{500}$.
Here we shall argue that this number is not at all ``large": it is
in fact perilously \emph{small}.

Leaving this claim to one side for the moment, having a ``large"
number like 10$^{500}$ emerge naturally from a physical theory is
of the utmost interest. For Nature herself has recently seen fit
to present us with an exceedingly \emph{small} number. Since 1998
\cite{kn:riess}\cite{kn:perlmutter}, evidence has been mounting
that an invisible negative-pressure energy component is currently
dominating the evolution of our Universe. The evidence for this
``dark energy" is strictly gravitational in nature, so, in stating
its properties, we use units appropriate to gravitation. These are
the Planck Units. An example of such units is the Planck density,
given in terms of ordinary units by $5.15500 \times 10^{96}$
kilogrammes per cubic metre. The dark energy mass density is given
in these units by
\begin{equation}\label{eq:B}
\rho_{DE}\;\approx\;10^{-\,120}.
\end{equation}
The first point to note with regard to this number is that all
attempts to compute a unique value for it, or for the related
quantity known as the cosmological constant, have failed. In
string theory, one has a likely candidate, the vacuum energy
density, but the theory does not fix this quantity
\emph{uniquely}: it varies across the Landscape. The second point
to note is that this number does appear to be very small indeed;
but perhaps it is not so small when we discover that we may have
something like 10$^{500}$ opportunities to realise it. This may
possibly be the case if we identify the observed dark energy with
the Landscape vacuum energy.

The situation is analogous to the following one. Suppose that you
have made it your life's goal to throw a dart at a dartboard in
such a manner that the dart strikes the ``bullseye". There are
various ways in which this ambition might be realised. You might
cheat, using perhaps a system of powerful concealed magnets to
ensure the desired outcome. A marginally less ignominious scheme
would be to throw handfuls of darts at the board, without
bothering to aim; with a few thousand darts, a bullseye will
probably be scored; the superfluous darts can then be removed.

Now suppose that a game official enters the room and finds a
single dart in the bullseye. The question as to which of our
suggested techniques has actually been used can only be settled by
examining the circumstances. If, for example, a scrupulous search
fails to reveal any magnetic devices, but does uncover a crate
containing several thousand darts, then it is reasonable to assume
that the second method has been employed.

Our failure to find a means of uniquely deriving a very tiny
cosmological constant from first principles, combined with the
existence of a huge number of string vacua, inclines many to
favour an account of the situation analogous to the second method
of dart-playing. That is, given a ``large" number of possible
vacua with suitably spaced vacuum energies, one can hope to show
that at least one string vacuum with the observed energy will be
internally consistent. (It should be said that actually making
this proposal work at a technical level is ---$\,$ contrary to a
widespread misapprehension
---$\,$ an extremely non-trivial and delicate matter: see \cite{kn:TASI}.)

Now the reader is asked to note the care with which we have chosen
our words: we say that the vacuum is \emph{internally consistent},
not that it \emph{exists}\footnote{There is of course a body of
philosophical opinion which controverts this distinction. Few
physicists sympathise; see however \cite{kn:tegmark}.}. There is a
very clear moral distinction between \emph{conspiring} to cheat as
a way of winning a professional darts game, and actually carrying
out such a conspiracy. More than this: if one actually tries to
score a bullseye by throwing thousands of darts at the board, one
needs to explain in detail exactly how one proposes to secure such
a large supply of darts.

Similarly, we need a way of actually building the universe whose
internal consistency we have asserted. The simplest approach is to
find some mechanism that automatically builds essentially
\emph{every} universe in the Landscape. This automatically brings
into existence universes with the ``right" value of the vacuum
energy.

The currently favoured mechanism for building universes is the
nucleation of ``baby universes," which split off from a given
universe under rather general conditions. Baby universes can have
properties which differ radically from those of their parents.
There are grounds for believing that this variability allows
essentially all of the Landscape to be brought into actual
existence, with baby universes having suitably spaced vacuum
energies, as above \cite{kn:TASI}. The idea is that baby universes
can themselves eventually have babies; thus, even if a particular
universe in the Landscape does not have a particularly small
vacuum energy, one or more of its \emph{descendants} may well have
a cosmological constant similar in value to the one we observe.

It seems to be generally assumed that, having found a mechanism
for producing the right value of the vacuum energy, our task of
universe-building is essentially complete; for even after
discarding babies with the ``wrong" vacuum energy, one still finds
an enormous number of babies which do resemble our Universe in
that respect. Surely a universe just like ours can be found in
this huge ensemble? Let us examine this assumption more closely,
however.

Isaac Newton discovered a set of laws which are able to describe,
with very great accuracy, the extremely complex motions of all of
the objects in our solar system. Yet the laws themselves are
extremely simple. How was that possible? It is possible because
Newton was able to isolate most of the complexity in the
\emph{initial conditions} of the problem. These initial conditions
are apparently not constrained by any law, and hence they can vary
in a very complex way. They have to be determined by direct
observation. Newton assures us that he can predict the motion of
the system \emph{provided} that we supply him with precise details
of the positions and velocities of the objects at some initial
time. Conversely, no matter how thoroughly we understand the
dynamical laws of a system, we will not be able to describe its
motion unless we are provided with the initial conditions.

Similarly here: even if we feel confident that our ensemble of
universes contains members with the ``right" physical laws and
parameters (such as the vacuum energy), we cannot say whether the
ensemble is likely to contain \emph{even one} member resembling
our Universe until we can answer this question: how hard is it to
produce a baby universe with initial conditions not too dissimilar
to those of our Universe?

The answer to this question is (roughly) known: it was given by
Roger Penrose \cite{kn:penrose} in 1979. Penrose observed that the
second law of thermodynamics, the statement that the entropy of an
isolated system can (almost) never decrease, implies that the
entropy of the Universe at the beginning of time must have been
very low, compared with the value it might have had. One can make
this more quantitative after the manner familiar in statistical
mechanics, as follows.

The basic concept we need is that of the ``phase space" of a
dynamical system. This is just an abstract mathematical space,
introduced (in this context) over a century ago by Josiah Willard
Gibbs, in which each point represents one possible state of the
system. In practice, many states are indistinguishable, so one
speaks of the system occupying a certain volume in phase space. To
say that the entropy of a system is low is to say that it occupies
a volume in the appropriate phase space which is small compared to
the full volume accessible to it. The volume in phase space
corresponding to the actual initial conditions of our observable
Universe can be roughly estimated, and Penrose computed the full
volume, using the then recently developed theory of the
thermodynamics of black holes. How much larger is this full phase
space volume, the volume that was open to the earliest Universe,
of which it somehow failed to take advantage? Penrose computes the
ratio to be about
\begin{equation}\label{eq:C}
P\;\approx \; 10^{10^{123}}.
\end{equation}
This number represents some fundamental property of our Universe;
we propose to name it the ``Penrose number". Note that the
exponent in the Penrose number is itself exponential. Compared to
\emph{this} number, the Landscape begins to seem positively
claustrophobic. Dividing Penrose's number by 10$^{500}$ has, to an
excellent approximation, no effect whatever.

The answer to our question ---$\,$ how hard can it be to build a
baby universe with initial conditions similar to the initial
conditions of our Universe?
---$\,$ has turned out to be: it is almost unimaginably hard.
\emph{If} we should propose to rely on ``mere chance" to select
for us a universe from the Landscape, then our chances of
obtaining the right value for the vacuum energy are excellent, but
the probability that this universe will resemble our own is
nevertheless \emph{essentially zero}. For it will almost certainly
have the wrong initial conditions.

The conclusion, of course, is that we \emph{cannot} rely on mere
chance to build anything that resembles our Universe. The
Landscape is far, far too \emph{small} for anything of that sort
to be possible. We commend this fact to the attention of those who
maintain, on the basis that 10$^{500}$ is ``large", that the
Landscape can account for \emph{any} possible observation.

There are many possible responses to this line of argument, and we
will discuss some of them below. But the essential point is clear:
no ``mechanism for building universes" can be considered
satisfactory unless it addresses the size of the Penrose number in
a convincing way.

As is well known, Penrose's calculation arose from his abiding
concern with the problem of the \emph{Arrow of Time}. The
macroscopic features of our world are massively asymmetric in
time. Eggs break when dropped; this process has never been seen to
occur in reverse. Since the seminal work of Ludwig Boltzmann, we
have understood why this is so: in modern terms, the intact egg
occupies a vastly smaller domain in phase space than the broken
one. That is, being broken is a more \emph{generic} state than
being intact. The breaking of the egg is then the entirely
unsurprising process of the egg finding its way from less to more
generic states. This evolution from less generic to more generic
is precisely what we call the passage of time. (A discussion of
this argument, and of the many subtleties we have ignored here,
may be found in reference \cite{kn:lebowitz}.)

The amazing aspect of this situation, then, is not the
irreversible breaking of the egg
---$\,$ it is the fact that intact eggs ever exist ``in the first place".
With this basic insight in hand, the existence of the Arrow that
we observe \emph{now} is readily traced back, through a chain of
steadily less generic states, to the earliest Universe. (The egg
owes its existence to a chicken, which in turn exists only because
the Sun supplies us with a reliable source of low-entropy photons,
and the Sun exists because of low-entropy conditions billions of
years ago, and so on.) The real question is then: \emph{why} was
our Universe born in such an ultra-low entropy state? What is the
ultimate origin of the Arrow of Time?

One point we wish to make here is that the string Landscape can
only be made to work if we can find a concrete answer to this
question. The need to answer it has suddenly become a matter of
the greatest urgency. If we cannot understand the Arrow, then we
will never know whether the Landscape has any physical relevance
at all; for we will not know whether \emph{any} universe in the
Landscape has the right initial conditions.

In this work, we will propose a solution of this problem. It is
put forward not as the last word on the subject, but to show that
the
---$\,$ undeniably formidable ---$\,$ difficulties which arise can
be surmounted. The basic idea is to combine the initial-value
theory for gravity with certain deep theorems in global
differential geometry to explain precisely how the Arrow came into
existence.

Our approach here will be resolutely string-theoretical. The
problem of the Arrow of Time can of course be posed in a broader
context, and there is a large literature on the subject from
various points of view. The classic text, giving a complete
treatment of the background material, is that of Heinz-Dieter Zeh
\cite{kn:zeh}; other indispensable sources are \cite{kn:albrecht},
\cite{kn:price}, \cite{kn:dyson}, \cite{kn:carroll}, and
\cite{kn:wald}. The new ideas advanced here are described in
technical detail in \cite{kn:arrow}\cite{kn:baby}; some recent
alternative approaches are given in
\cite{kn:kiefer}\cite{kn:laura}.

Let us begin with a survey of the difficulties which must be
overcome by any theory of the Arrow.

\addtocounter{section}{1}
\section*{\large{\textsf{2. Why Is Understanding The Arrow A Hard Problem?}}}
As befits a question of such fundamental importance, understanding
the origin of the Arrow is an extremely difficult problem. Let us
try to be as specific as possible as to \emph{why} it is so hard.

\subsubsection*{{\textsf{2.1. Inflation Doesn't Make an Arrow.}}}
\emph{Inflation} is the extremely successful idea
\cite{kn:lindereview} that our Universe is so large because of a
short but fantastically rapid period of expansion very early in
the history of the Universe. In this picture, the observed
Universe (and, in fact, a vast region beyond it) is obtained, by
the passage of time, from a tiny region near the beginning. This
theory has its critics, but it is fair to say that it is now the
standard theoretical framework for the physics of the very early
Universe. Basically, when one has a problem regarding the physics
of that era, one looks to Inflation for a solution\footnote{The
reader is nevertheless warned that the precise way in which
Inflation arises in string theory is not yet fully understood; see
\cite{kn:tegmark2}.}. One reason for the difficulty of the Arrow
question is that this is one of the few problems that Inflation
does \emph{not} solve. Let us see why this is so.

\emph{Unitarity} can be described informally ---$\,$
mathematically precise definitions can be given both in classical
and in quantum mechanics
---$\,$ as the principle that information is never completely lost
(or created \emph{ex nihilo}) in the evolution of any physical
system. All of the ordinary observed\footnote{Here and henceforth,
we use the term ``observer" in the metaphorical sense traditional
in relativity, that is, to mean a timelike curve, usually a
geodesic. \emph{Absolutely no ``anthropic" connotations are
intended}; ``observers" in our sense exist at all times in every
universe.} processes in nature obey this principle; which is not
to say that it cannot, in principle, be violated by very exotic
processes.

Now it does not follow that a description or portrait of an object
must have as many possible states as the object itself. An atlas
of maps of the Earth certainly has a smaller range of possible
states than the Earth itself. This is of course by no means
paradoxical, since it is clearly understood that the art of making
maps consists in deliberately disregarding irrelevant information
in order to present more important data; that is, cartography is a
process in which information is discarded, not destroyed. To many
physicists, it seems ``natural" that the inflationary ``initial
region" contained very little information: after all, it was
extremely small, whereas the current Universe is enormously large;
one is tempted to think of the early universe as a sort of plan or
map. But
---$\,$ Sean Carroll and Jennifer Chen \cite{kn:carroll} have stressed this simple
yet fundamental point ---$\,$ the early Universe was not a
\emph{map} of the present world: it \emph{was} that world. The
latter is obtained from the former simply by means of the passage
of time, which is ordinarily taken to be a unitary process. If
unitarity holds, then the smallness of the initial region
\emph{does not} explain why its entropy was so low.

Now one can certainly argue that the evolution of the Universe is
no ordinary process; perhaps it is precisely an ``exotic" process
which does violate unitarity, admittedly on a massive scale.
Indeed, as is well known, the ``exotic" process of the evaporation
of black holes was long widely believed to violate unitarity
\cite{kn:hawkingevap}, and the arguments in favour of that belief
were very cogent. This was not considered to be contradictory,
because black hole evaporation is certainly outside the range of
our ordinary (or other) experience.

Despite this, it is rapidly becoming the consensus view that black
hole evaporation \emph{does not} violate unitarity. This
unexpected development is due to the discovery of ``AdS/CFT
duality" in string theory. Very briefly, this is the idea that
quantum-gravitational processes, such as black hole evaporation,
can be completely described by a certain (``dual") quantum field
theory. Since this latter is fully unitary, so also must be the
physics governing the equivalent evaporating black hole
\cite{kn:juan}.

This development has convinced many that unitarity is always
strictly preserved in string theory. But if one accepts this, it
follows, as we have seen, that the mere fact that our Universe has
inflated \emph{does not} explain the Arrow of Time in the context
of string theory. Once again, we see that a recent development in
string theory has greatly increased the urgency of solving the
problem of the Arrow. For if we believe in unitarity, then the
initial region must have had the potential to acquire an amount of
entropy that is stupefyingly large compared to its size. The
situation is analogous to opening an atlas of world maps, putting
a certain map under a microscope, and seeing oneself represented
accurately on the page. Indeed, it is far more astonishing than
that.

In short, Inflation means that a recent development in string
theory
---$\,$ the confirmation of the universality of unitarity ---$\,$
only really makes sense if we can solve the problem of the Arrow.
Conversely, if we wish to solve that problem within string theory,
we have to do it \emph{without} appealing to Inflation.

In fact, the situation is exactly the reverse: Inflation depends
on low-entropy initial conditions in order to get started. To see
this, recall that Inflation is driven by a scalar field, the
``inflaton", denoted $\varphi$. \emph{Generically}, such a field
depends strongly on position both in time and in space: that is,
generically, its space and time derivatives,
$\partial_{\mu}\,\varphi$ (where $\mu$ is an index labelling the
three space directions, plus time), will not be close to zero. The
distribution of energy and momentum in the inflaton field is given
by its stress-energy-momentum tensor $T_{\mu\nu}$, which takes the
form
\begin{equation}\label{eq:D}
T_{\mu\nu}\;=\;\partial_{\mu}\varphi\;\partial_{\nu}\varphi\;-\;
{{1}\over{2}}\,g_{\mu\nu}\,(g^{\alpha\beta}\,\partial_{\alpha}\varphi\;\partial_{\beta}\varphi\;-\;V(\varphi)).
\end{equation}
Here $g_{\mu\nu}$ is the metric tensor familiar from General
Relativity, there is a summation implied over the indices $\alpha$
and $\beta$, and $V(\varphi)$ is a scalar function of $\varphi$
called the inflaton potential. We need not discuss the details: it
is intuitively clear from this expression that the energy stored
in $\varphi$ can be distributed in a vast variety of ways. But for
Inflation to begin, the energy must be distributed in an
\emph{extremely special} way; Inflation will only occur if, to an
excellent approximation, all of the derivatives of $\varphi$ are
initially zero, and the potential is positive. For then we have
\begin{equation}\label{eq:E}
T_{\mu\nu}\;\approx\; {{1}\over{2}}\,g_{\mu\nu}\,V(\varphi),
\end{equation}
and it is well known that energies of this kind lead to a rapidly
accelerated expansion (since then the potential behaves like a
positive cosmological constant). One says, in view of
(\ref{eq:E}), that Inflation begins with the inflaton in its
``potential-dominated state" \cite{kn:albrecht}. The point is that
the potential-dominated state is clearly an extremely non-generic
state for the inflaton: that is, Inflation can only begin if the
inflaton has somehow been put into this very low-entropy state.
Inflation depends on the existence of an Arrow of Time, which must
have been established \emph{before} Inflation began.

This should not be construed as a criticism of Inflation. On the
contrary, Inflation gives us an enormous simplification of the
problem of explaining the Arrow. As Huw Price has emphasised
\cite{kn:price}, we are by no means entitled \emph{a priori} to
assume that the low initial entropy of our Universe was ``stored"
in a simple way or confined to a single form of matter. Inflation
performs precisely this service: it implies that the entire,
potentially vastly complicated problem of accounting for the
entropy of the early Universe reduces to the single problem of
accounting for the initial state of the inflaton. In the words of
Lisa Dyson, Matthew Kleban, and Leonard Susskind \cite{kn:dyson}:
``Some unknown agent initially started the inflaton high up on its
potential, and the rest is history." In fact, at stake here is the
reason why there \emph{was} any history: if we can understand the
``unknown agent", then, thanks to Inflation, we will have a theory
of the ``passing" of time.

In short: the first reason for the extreme difficulty of our
problem is that while Inflation is undoubtedly a crucial part of
the puzzle ---$\,$ later we shall see that it is even more
important than is apparent from this discussion ---$\,$ it
\emph{assumes} a pre-existing Arrow; so we are deprived of a major
technical resource.

\subsubsection*{{\textsf{2.2. Laws of Nature vs Initial Conditions}}}
To say that the very early Universe had ``low" entropy is not to
say much; this vagueness, indeed, is one of the obstacles to
understanding the origin of the Arrow. Fortunately, as we have
already discussed, Penrose has solved this problem for us, in
terms of the fundamental number $P$ (equation (\ref{eq:C})). We
are searching for an explanation of ``low" entropy, where ``low"
is quantified by the fantastically small number $1/P$.

However, the extreme smallness of this number is itself another
factor in the difficulty of explaining the Arrow. We have seen
that string theory can naturally give rise to ``large" numbers,
such as $10^{500}$, and the reciprocal of this is, by any normal
standard, an extremely small number. But it is still vastly too
large compared to $1/P$. What kind of theory could possibly
generate numbers of this order?

One possible answer to this question can be stated in simple
language as follows: in an eternal universe, \emph{everything} can
happen, even something with probability of order $1/P$. Of course,
something so improbable will happen rarely, but what of that?
\emph{Just wait}.

This ``just wait" argument goes back to Boltzmann himself
(\cite{kn:boltzmann}; see \cite{kn:price} for a clear discussion
of this theory). The idea is simply that the extremely low entropy
of the early Universe was due to a random fluctuation, such as
must always occur in any system at equilibrium, given enough time.
The subsequent history of the Universe is just the normal return
to the prior equilibrium state. In essence, Boltzmann and his
successors claim to solve the problem of the Arrow by outbidding
Penrose: to his number $P$ they respond with a (supposedly)
\emph{infinite} supply of time. (There is also a spatial version
of this argument: in a spatially infinite universe, everything, no
matter how improbable, will happen somewhere. All of our
strictures on the ``just wait" argument apply, mutatis mutandis,
to this version.)

This idea, with its air of ``getting something for nothing", has a
strong superficial appeal; but its logical status is, in fact,
very problematic. It is certainly far from clear, for example,
that \emph{everything} can happen even with unlimited resources of
time or space. More seriously: why stop at ``explaining" the
initial conditions of the universe in this way? The Universe we
observe has many other ``improbable" features; notably, it has
\textit{laws of nature}; why not ``explain" these, too, as
\emph{apparent} regularities such as must arise, by mere chance,
if one is prepared to wait sufficiently long? It is possible that
Boltzmann's ``rare fluctuation" can account for the low entropy at
the beginning of time; but it is also \textit{possible} that the
Universe fluctuated into existence last Thursday
\cite{kn:thursday}, complete with suitable false evidence of being
ancient. The mere fact that an explanation ``works" does not in
any way oblige us to take it seriously. One suspects that the
``just wait" technique is applied exclusively to events in the
remote past or future for a reason: the phenomena are not present
to reproach us with the extreme implausibility of this kind of
argument.

Physical systems typically have characteristic time (and length)
scales. One does not expect plate tectonics to predict the motions
of continents on microsecond time scales or nanometre length
scales. Similarly, the ``just wait" philosophy can make sense only
in the context of a theory which incorporates (possibly vast but)
\textit{finite} characteristic time and space scales. For these
scales will impose discipline on the ``just wait" philosophy;
requests for more time or space than the characteristic scales can
be denied (unless one has a mechanism explaining the unusual
longevity of a specific system, which is not the case here).

Indeed, string theory does seem to have such scales. The known
ways of constructing string vacua resembling our Universe
\cite{kn:KKLT} (in that they have a \emph{positive} cosmological
constant) certainly \emph{do not} lead to universes that endure
for arbitrarily long periods of time; one says that the vacuum is
``metastable". One might wait uncounted aeons for Boltzmann's
``rare fluctuation" to occur, only to find that the system being
examined has by that time disappeared into baby universes which
nucleated in the meantime\footnote{One will always be able to find
particular ``fortunate" systems which evade this fate, since the
babies do not devour the entire space \cite{kn:carroll}.
Sufficiently fortunate systems can endure far beyond the
characteristic time scale. Such an argument cannot be disproved;
which is precisely why it should be rejected.}. Similarly, the
question as to whether the Universe is spatially infinite becomes
somewhat ambiguous in string theory: the ``holographic principle"
(see for example \cite{kn:bouff} and references) apparently
indicates that the regions beyond a cosmological
horizon\footnote{A cosmological horizon is a surface surrounding a
cosmological observer, such that objects beyond it are forever
invisible to that observer. Such horizons arise in cosmological
models, such as de Sitter spacetime, in which a positive vacuum
energy dominates all other forms of matter.} are in some sense
merely a \emph{redundant description} of the data inside the
horizon. This may well imply that something that is very unlikely
in one cosmological domain is in fact unlikely to occur anywhere.
In short, it is far from clear that, in reality, we do have the
limitless resources of time and space postulated by the ``just
wait" theories. String theory does allow for periods of time and
regions of space which are truly vast by any normal standard.
Boltzmann-style fluctuations might be legitimate objects of
scientific interest if they occurred within those allowances;
this, however, is apparently \textit{not} the case.

In fact, Boltzmann's argument is usually held to fail
\cite{kn:price} because of reasons connected with the ``anthropic
principle", the idea that the existence of intelligent life
imposes non-trivial conditions on physical theories. However, this
principle has itself been strongly criticised
\cite{kn:wald}\cite{kn:krauss}. In our view, it is neither
necessary nor desirable to become involved in that highly
contentious debate here. In fact, we feel that the entire ``just
wait" approach is misconceived, both because of the remarks above
and because of the following argument.

Normally, when we discover some ``highly improbable" pattern in
Nature, our reaction is that the seeming improbability is really a
sign of the existence of some hitherto unknown \emph{law of
Nature}. When we discover, for example, that planetary orbits have
the approximate shape of an ellipse ---$\,$ a shape which has been
known to be of mathematical interest for thousands of years
---$\,$ nobody would think it an adequate response to point out that
such paths are bound to be traced out by randomly moving particles
in the fullness of time. Instead we search for a law, or a
collection of laws (in this case, Newton's laws of motion and of
gravitation). When these laws are understood, the improbable is
transformed into a certainty; elliptical orbits are an unavoidable
consequence of the laws. Similarly here: surely the normal
scientific response to Penrose's observation is that the initial
conditions of our Universe seem improbable only because we do not
know the underlying law of Nature.

In our view, Penrose's calculation really means that the initial
entropy was as low as it is mathematically possible for it to be:
in other words, ``no greater than $1/P$" is really just a way of
saying ``exactly zero at the classical level". Again, this is the
kind of statement which would result from the existence of some
law of Nature governing the initial conditions of the Universe.

Another way of stating the case is as follows. One might suppose
that the low entropy conditions at the beginning of time may have
been selected by some kind of quantum-mechanical probability
distribution (defined on the ``set of all initial conditions").
The problem here is, once again, the sheer size of the Penrose
number; somehow we have to find a way of concentrating the
wave-function over a particular outcome, and we must concentrate
it to an unheard-of degree. This suggests that we are dealing with
a quantum-mechanical version of a classical situation in which
only a truly \emph{unique} configuration is possible. (That is,
the classical probability distribution is a delta function.) But
this is tantamount to saying that we must find a (classical) law
of Nature which dictates the nature of this unique set of possible
initial conditions.

This point of view has the following major advantage. One of the
objections (discussed, for example, in reference \cite{kn:davies})
to Boltzmann's explanation of the second law of thermodynamics is
the following. We said that a system has low entropy if it is
confined to a small volume in phase space. This volume is occupied
by a set of states which are \emph{indistinguishable} from each
other. But indistinguishability is to some extent a ``subjective"
matter: it has something to do with the power of the observer to
distinguish. (This process of dividing up the phase space into
sets of indistinguishable ``micro-states" is known as
``coarse-graining"; see \cite{kn:albrecht}.) By requiring the
initial state to be \emph{absolutely unique} at the classical
level, the law of Nature we seek helps to resolve this problem.

This imposition of a \emph{law} on initial conditions puts us,
however, in a very unfamiliar situation. As we discussed in the
Introduction, it was one of Newton's great insights that extremely
complex behaviour can be understood in terms of very simple laws
precisely by rigorously separating laws from initial conditions.
Thus, another reason for the difficulty of our problem is that it
requires us to find a way of breaking down this traditional
separation.

Of course, one could simply declare that the new law is: ``entropy
must be as low as possible at the beginning of time". This is not
satisfactory because it is not related to, and does not flow from,
our other theories of the early Universe: Inflation and string
theory. The new law should be connected with Inflation; and it
should, ideally, be an inescapable consequence of the mathematical
structure of string theory. But these statements do not make the
problem any easier.

\subsubsection*{{\textsf{2.3. Low Entropy is a Geometric Property}}}
Thus far, we have spoken of ``low entropy" in a rather vague way.
In order to progress, we need to refine this expression. The way
to do this was pointed out, once again, by Penrose
\cite{kn:penrose}.

We begin by noting that, by the time the Universe had evolved to
the point where its contents are directly observable by us (that
is, the stage called ``decoupling"), most of these contents are by
no means in a low entropy state. On the contrary, the cosmic
microwave background has nearly all of the characteristics of a
\emph{high} entropy state: its spectrum is that of a black body.
The only only hint of the extreme ``specialness" that defines a
low-entropy state is the extreme spatial uniformity of the
radiation: to a good approximation, it looks the same in all
directions. One interprets this as a sign of low
\emph{gravitational} entropy, as follows.

The precise way in which we should compute the entropy associated
with a gravitational field is not yet fully understood, but the
existence of black hole entropy (associated with its horizon, see
for example \cite{kn:ross}) assures us that the concept does make
sense. Now the natural tendency of gravitational systems (except
in circumstances which do not hold here
---$\,$ see section 3.2 below) is to become clumped, indicating that
gravitational entropy is somehow related to the ``degree of
clumping". But the cosmic background radiation, despite its
thermal character, is extremely uniformly distributed: it is
hardly clumped at all. Thus, in ``looking" at the microwave
background, we are observing a system in which the gravitational
entropy, and \emph{only} the gravitational entropy, is low. This
very strongly suggests that ``low entropy" in the very earliest
Universe, even when we cannot observe it directly, really means
``low gravitational entropy."

Now General Relativity teaches us that gravitation is nothing but
a manifestation of the geometry of spacetime. Therefore, low
entropy for gravitation must correspond to the spacetime geometry
taking some very special form; ``specialness" here actually means
extreme spatial \emph{uniformity}. This refines our problem very
substantially: the problem of explaining the low entropy of the
early Universe has been reduced to explaining its spatial
uniformity.

This point is of such fundamental importance that we should expand
on it. Think of a two-dimensional sphere. It has a characteristic
\emph{topology}, expressed in this simple case by saying that it
is compact and has genus zero (it has no ``holes"). It also has a
distinguished \emph{geometry} associated with a perfectly round
shape. But clearly there are infinitely many other geometries
compatible with this topology: one can deform the two-sphere to an
arbitrarily complex shape without tearing any holes in it. The
perfectly round shape is obviously very ``special"; a typical
geometry, chosen randomly from the space of all possible shapes
with spherical topology, will look nothing like this. Furthermore,
two such typical shapes will be hard to distinguish, just as it
would be difficult to say whether two clouds resemble each other.
In the language we have been using: the perfectly round spherical
geometry corresponds to low gravitational entropy. Similar
comments apply to other spaces. For example, the familiar flat
geometry described by Cartesian coordinates is the geometry with
the ``lowest possible entropy" on a space with that topology.

The general situation is as follows. For a topological space with
a given number of dimensions, it turns out that there is a measure
of the amount of symmetry that a given geometry can have in a
neighbourhood of any point\footnote{Technically, this measure is
the maximal dimensionality of the isometry groups of the metrics
induced on open sets containing the point.}. There is a maximum
possible ``amount of symmetry" in each dimension, and certain
topologies allow this maximum to be realised by particular
geometries; the topologies of interest to us here are of this
kind. Penrose's calculations indicate that the earliest spatial
sections of our spacetime were, to a precision given \emph{at
worst} by the reciprocal of Penrose's number, similar to some such
maximally symmetric three-dimensional space.

This actually ties in rather beautifully with our earlier
discussion of Inflation. We saw that the latter depends on the
existence of some ``unknown agent" which is capable of putting the
inflaton into its potential-dominated state initially. Now the
extremely symmetric geometries we are dealing with here have the
property of being everywhere locally isotropic: that is, the
geometry ``looks the same in all directions" to an arbitrarily
situated observer\footnote{Except in certain very special
spacetimes, isotropy can only be seen by a unique family of
observers, who are distinguished by that very fact.} (who cannot
however see arbitrarily far
---$\,$ hence the ``locally"). Now suppose that we can show that the
inflaton field must share the symmetries of the underlying
three-dimensional space. Energy cannot flow under such conditions,
because the direction of flow would define a ``special" local
direction, which is precisely what local isotropy forbids. In
fact, under these conditions, physical fields are forbidden to
define \emph{any} special direction. But this is a formidable
restriction indeed, because most physical fields are defined, at a
fundamental level, by \emph{vectors}, which, by their very nature,
do define a direction. In fact, the only way a vector can be
isotropic is by being exactly zero. But the spatial derivatives of
the inflaton define a vector. Thus, if the field shares the
symmetries of the underlying space, a perfectly locally isotropic
spatial geometry forces the spatial derivatives of the inflaton to
vanish.

This goes very far ---$\,$ though we still have to explain the
smallness of the initial \emph{time} derivative of the inflaton
---$\,$ towards putting the inflaton in its lowest-entropy state,
just as we wanted. More than that: this argument strongly suggests
that the earliest form of matter probably \emph{was} a scalar
field, precisely as Inflation postulates. For a scalar field can,
unlike a fundamental vector field, be non-zero even in conditions
of perfect isotropy\footnote{A similar argument applies to spinor
fields. Note that \emph{tensor} fields can be non-trivial even
when they are perfectly isotropic; it is the fact that general
relativity represents the gravitational field by a tensor which
permits non-trivial isotropic cosmologies.}. In short, extreme
spatial symmetry, interpreted as local isotropy around every
point, is an ideal candidate for the ``unknown agent" which
started Inflation, and the argument we are developing here is,
more generally, extremely natural from the inflationary point of
view.

In the previous section we asked: what kind of theory can give
rise to ``low" entropy on the kind of scale measured by $1/P$? Our
discussion here reduces this to: what kind of theory can give rise
to geometric \emph{symmetry} (more precisely: local isotropy at
each point) with such fantastic accuracy? Notice that, by
sharpening the question in this way, we are drastically raising
the bar for any proposed explanation. A theory which ``merely"
predicts ultra-low entropy is not good enough; it has to predict
low entropy of a specific, \emph{geometric}, form.

A good example of what we have in mind here is provided by one of
the most carefully thought-out proposals for explaining the Arrow,
due to Carroll and Chen \cite{kn:carroll}. In this theory, baby
universes (of a particular kind, see below) branch off from a
mother universe by means of a rare fluctuation in the inflaton
field, a fluctuation \emph{upwards} to a higher-energy
equilibrium. The basic idea is that the mother is in a \emph{high}
entropy state (which is generic, hence not itself in need of
explanation) but the baby is in a \emph{low} entropy state. As one
would expect from this, however, this theory demands Boltzmannian
patience\footnote{Carroll and Chen estimate the relevant
probability to be $\approx\;10^{\,-\,10^{10^{56}}}$, a figure
which induces awe and presbyopia in equal measure.} as we await
the nucleation of a baby universe with sufficiently low entropy;
that is, the Carroll-Chen theory is a ``just wait" theory
\emph{par excellence}. In this sense, in fact, the theory is much
like Boltzmann's. The difference lies in the claim that
Boltzmann's scenario arises naturally as a consequence of the
modern understanding of early-universe physics and spacetime
dynamics.

One way to explain how this theory works (that is, why the upward
fluctuation of the inflaton produces a low-entropy state) is as
follows. Let us assume that the spacetimes of both mother and
child can be approximated by de Sitter spacetime, but with a much
larger (inflationary) value of the cosmological constant for the
baby than for the parent. Now, like the horizon of a black hole,
the cosmological horizon of de Sitter spacetime has an entropy
associated with it. Large values of the cosmological constant
correspond to \emph{lower} entropy, so this is consistent with the
claim that the mother has much higher entropy than the baby. Thus
the theory ``explains" why the inflaton was in a low-entropy state
initially.

This argument is, however, \emph{circular}. The usual formula
\cite{kn:gibhawk} for de Sitter entropy (in terms of the horizon
area) is of course derived using de Sitter spacetime geometry,
which is maximally spatially symmetric \emph{by assumption}. If
the baby's spatial sections were, on the contrary, extremely
anisotropic, then we would have no grounds for using the de Sitter
entropy formula to estimate its entropy (which in that case would
in fact not be low, or at least not low enough). Thus, the
Carroll-Chen proposal will only work if one can show that the baby
is born with very isotropic spatial sections. The above
calculation of the baby's entropy in terms of the value of its
inflationary potential is certainly \emph{consistent} with a
possible solution of our problem; but in itself it does not
provide an explanation, because it does not address the fact that
``low entropy" really means ``extreme initial spatial isotropy
around every point".

In fact, if the mother universe has high entropy, this means that
its spatial sections are \emph{not} highly isotropic around each
point; they must be highly irregular (and hence not like those of
de Sitter spacetime); and yet the mother must give birth to
extremely isotropic babies. Explaining how that can happen is
precisely the difficulty.

We will see later that the Carroll-Chen theory can in fact be
interpreted so that it deals with this problem. For the moment,
the point we are making is that there is little hope of explaining
the Arrow unless we can address the problem of spatial regularity
\emph{directly}. The remainder of the explanation will then
follow.

Our emphasis on isotropy is motivated by Penrose's ``Weyl
Curvature Hypothesis" \cite{kn:penrose}. Penrose postulates that
the initial geometric ``specialness" takes the form of a demand
that a certain piece of the spacetime curvature tensor, the Weyl
tensor, should vanish at initial singularities (and \emph{only}
there\footnote{A more precise technical formulation is that
spacetime should be ``conformally compactifiable" at initial
singularities.}). This demand does ensure local isotropy around
each point ``in" the initial singularity. However, there is a
serious problem here: most physicists hope that the presence of
singularities in the standard cosmological models is misleading. A
full understanding of the initial state should reveal it to be
\emph{non-singular}. How can Penrose's idea be adapted to this
case?

In our view, local isotropy around every point is in any case the
heart of the matter, and one might as well try to impose it
directly. This has the great advantage that it applies also to the
case of an entirely \emph{non-singular} cosmology. But it also
forces us to confront a central technical difficulty.

Once we accept that the origin of the Arrow is reducible to
explaining the ``special" geometry of the earliest spatial
sections of the Universe, we can frame the problem in the language
of statistical mechanics as follows. Consider the set of all
possible three-dimensional Riemannian manifolds. Think of this as
a ``phase space". In this phase space, the subset of manifolds
which are everywhere locally isotropic is a fantastically tiny
subspace. It consists of familiar geometries like that of the
perfectly round three-dimensional sphere, or of perfectly flat
spaces, or of geometries like that of hyperbolic space, the space
having constant negative curvature. The generic three-dimensional
geometry is nothing like any of these, \emph{even approximately}:
one should picture it as a wildly distorted ``blob", corresponding
indeed to just what we mean when we say that something is
``shapeless". We stress that it is \emph{not} enough just to study
spaces with geometries which are obtained by making small
perturbations around the perfectly isotropic spaces. For even
these will sample an insignificant region in the full phase space.

To see more concretely what we mean by this, let us consider one
of the most sophisticated attempts to \emph{derive} an Arrow of
Time from a more basic formalism, the one due to Stephen Hawking
\cite{kn:hawking} and his collaborators \cite{kn:laf}. Here one
uses the well-known ``no boundary" version of quantum cosmology to
study inhomogeneous perturbations around an assumed perfectly
isotropic background. A subtle technical argument leads to the
conclusion that the perturbations are ``born" in their ground
state, and thereafter grow, defining an Arrow of the appropriate
(geometric) kind. This is undoubtedly an extremely important check
of the \emph{consistency} of the no-boundary approach with the
existence of an Arrow of the kind we observe. But in itself it
does not solve the main problem, since the perfect isotropy of the
background is \emph{assumed}.

The problem here is that the full phase space is truly vast and
very difficult to describe in detail
---$\,$ even in the case of two dimensions, it is hard to describe the ``generic"
shape into which a sphere can be deformed
---$\,$ and correspondingly ``hard to manage". By this we mean something specific:
there are very few mathematical theorems which operate at this
level of generality. A typical theorem in differential geometry
will begin with much more restrictive assumptions: as for example,
``consider the set of compact manifolds with curvature constrained
in the following manner..." Theorems of the form: ``Consider the
set of \emph{all} compact three-dimensional manifolds..." are rare
indeed. This set is simply too large and its contents too various
and hard to characterize.

Even if we had the relevant theorem, it could only lead us to a
conclusion if we were able to apply it to some well-justified
\emph{physical} assumptions. But it is very likely that the
conditions we can justify will be extremely weak. They will, for
example, pertain only to the spatial surface along which the
Universe came into being, not to the entire subsequent history.
They will not involve strong assumptions about the behaviour of
matter in the beginning, because we have no reason to believe that
matter behaved, at that time, in ways familiar to us now.

An example of the kind of argument we have in mind was advanced in
an important work of Gary Gibbons and James Hartle
\cite{kn:gibhart}, who attempted to show, using a powerful theorem
in global differential geometry, that the initial spatial section
of the Universe must have been a perfectly round three-dimensional
sphere. The argument does \textit{not} assume that the geometry
must be approximately spherical; it surveys the entire region of
the phase space compatible with its physical assumptions.
Unfortunately, those assumptions are much too strong, to the
extent that they are almost certainly not satisfied by the actual
Universe. (See \cite{kn:maldacena} for a detailed discussion of
these conditions.) Furthermore, in the context of Inflation, the
discussion in \cite{kn:gibhart} effectively assumes that the
inflaton has been put into its minimal-entropy
(potential-dominated) state; but we have argued that this state
must be a \textit{consequence} of the existence of an Arrow. So
the conditions needed here are far too restrictive.

Thus, we have another reason for the difficulty of our problem:
despite our efforts to refine it, it remains at a level of
generality such that technical resources are scant. We need a
technique which can take the \textit{very mild} physical
conditions that we are entitled to impose at the beginning of
time, and transform them into an enormously restrictive constraint
on the truly vast phase space of initial geometries. In addition,
it has to force the inflaton to share the symmetries of the
resulting geometry, and it will also have to restrict the way the
initial spatial slice is embedded in spacetime (so that it can
tell us something about the initial \emph{time} derivative of the
inflaton). The reader can be excused for doubting that such a
magical technique exists at all.

\subsubsection*{{\textsf{2.4. In My Beginning Is My End}}}
Thus far, we have spoken as if our sole task is to explain the
peculiar conditions which obtained at the \textit{creation} of a
universe. If we can perform this task, however, then we should be
able to understand the behaviour of the Arrow everywhere in
spacetime, including at the \textit{destruction} of a universe.
Indeed, an inability to answer such questions would be a strong
indication that something is wrong with our theory. Note that some
universes in the Landscape have \emph{negative} vacuum energy,
which does indeed cause them to destroy themselves; thus we cannot
evade this question by noting that \textit{our} Universe shows no
signs of mortality. Furthermore, even in our Universe, spacetime
is ``destroyed" inside black holes. We need to be able to predict
what the Arrow would be like for observers in less salubrious
circumstances than our own.

Most of us feel intuitively that \emph{initial} conditions
completely fix the fate of a universe, that its end is implicit in
(and different from) its beginning. That is, we feel that there
are no ``special conditions" when a universe is destroyed; we find
it hard to accept that the direction of the Arrow can somehow
``flip", as it would have to do if conditions near the destruction
of a universe are as special as those near its creation. Can this
intuition be justified
---$\,$ or is it no more than a prejudice? Certainly, the arguments
that we instinctively raise against a symmetry between creation
and destruction seem invariably to convict us of ``cosmic
hypocrisy": we find ourselves arguing about cosmic destruction in
ways which we would not dream of applying to the creation. It is
no use saying, for example, that low entropy in (what we call) the
future is ``highly improbable": that is true, but it is equally
true that low entropy in the \textit{past} is ``highly
improbable"; it is absurdly improbable, yet it was indeed so.
Price \cite{kn:price} has dissected many fallacious arguments of
this kind.

Despite this, we propose to argue that our intuitions are not
misleading us in this case: we will claim that there are indeed no
``special conditions" at the destruction of a universe. We can
begin to clarify these issues by eliminating one major source of
confusion: the expansion and (in some cases) contraction of
universes.

The observed thermodynamic Arrow runs parallel to the expansion of
the Universe. Since the former is generally agreed to have a
cosmological origin, one sometimes speaks also of a ``cosmological
Arrow". It is then natural to ask whether there is some kind of
link between the two. In fact, some authors ---$\,$ see
\cite{kn:davies} for a particularly eloquent defence of this idea
---$\,$ have gone to the extreme of arguing that the thermodynamic
Arrow is a \textit{consequence} of the cosmological Arrow.

It is important to understand that this idea \textit{cannot} be
correct if Boltzmann's interpretation of the passage of time,
accepted here and by most authors, is valid. For if the
thermodynamic Arrow depends on the cosmological Arrow, it follows
that the former could not exist if the Universe did not expand.
But recall that Boltzmann pointed out that the problem of the
Arrow reduces to explaining the special initial conditions of our
world; if that can be understood, then the subsequent evolution
towards more generic states is only to be expected, and it will
occur \textit{whether or not} the Universe expands. In fact,
granted that we can somehow explain the special initial
conditions, entropy will increase, at least initially, even if the
Universe always contracts.

If we accept Boltzmann's account of the basis of the second law,
then, it follows that the expansion and contraction of a universe
are irrelevant distractions as far as understanding the Arrow is
concerned. The real question is this. Consider a universe which is
both created and destroyed: is there any justification for
thinking that ``special" conditions, such as spatial uniformity,
can be imposed at one ``end" of the universe and not the other?
(While we argue that the expansion of the Universe is not
\emph{responsible} for the Arrow, it is of course still possible
that both the expansion and the Arrow have a common origin. This
is precisely what happens in the theory of the Arrow discussed
below and in \cite{kn:arrow}.)

If a universe has two ``ends", then the only difference between
the ends is precisely the difference, if there is one, between
\emph{creation} and \emph{destruction}. Now it \emph{may} be that
these two words have no meaning apart from the one given to them
by the existence of an Arrow: ``creation" may be another
\emph{name} for the low-entropy end, ``destruction" may be a
\emph{name} for the high-entropy end. But this is not at all
obvious. Why should the degree of ``specialness" of the state of
the universe have anything to do with its coming into existence?
How high does the entropy have to be before one is allowed to
destroy a universe? One can all too easily imagine a universe that
is created in a generic state, and remains in that state until it
is destroyed. The alternative claim, that such a universe is a
logical impossibility, seems far-fetched. Certainly the onus is on
those who assert this to prove it.

We shall assume, therefore, that ``creation" and ``destruction"
have some meanings which are logically prior to the Arrow; that
is, they might be meaningful even in universes which do not have
an Arrow. Nevertheless, if an Arrow does exist, it has to conform
itself to these meanings: it would clearly be unpleasant to find
oneself saying eventually that a certain universe ``began by being
destroyed". It should be possible to \emph{show} that entropy was
low at the creation, and will be high at the destruction. This
will have the \emph{consequence} that there are no ``special
conditions" at the destruction.

All this seems very reasonable. Once again, however, we are asking
for a great deal here. First, we must formulate the distinction
between creation and destruction in concrete mathematical
language. Then we must find a mechanism which imposes special
conditions on the ``creation end", \emph{and} which somehow fails
to do so at the ``destruction end" of a universe. It is one thing
to make ``reasonable" assertions about these matters, quite
another to establish (or even formulate) them mathematically. We
have yet another major obstacle to overcome.

\subsubsection*{{\textsf{2.5. Summary}}}
The difficulties we have discussed give rise to an intimidating
list:

\begin{itemize}
  \item First, we cannot rely on Inflation to help us; Inflation
  needs an Arrow in order to get started.
  \item Second, it seems unlikely that we will obtain an Arrow (of
  the kind we actually observe) simply by waiting, in the manner of
  a Boltzmann. Instead we need something more akin to a new law of
  Nature which, unlike any other, constrains the initial conditions
  of the Universe.
  \item Third, this law must somehow constrain the \textit{geometry} of the
  initial spatial section, by means of some technique which has the power to probe
  the entire vast space of all possible three-dimensional
  geometries.
  \item Finally, the theory should determine the behaviour of the
  Arrow of Time not just in the vicinity of the beginning of time,
  but throughout spacetime, including near its destruction. If we assert
  that one end of a universe is dissimilar to the other, we have to \emph{prove}
  this.
\end{itemize}

Having summarized the worst of our difficulties, we can now
propose an approach to solving them. We begin by discussing the
currently favoured way of making universes, the nucleation of
babies.

\addtocounter{section}{1}
\section*{\large{\textsf{3. Bringing Up Baby}}}
In the Landscape, a universe begins as the baby of some mother
universe. In this section we argue that the baby universe will not
spontaneously develop an Arrow; it can only inherit one from its
mother. If all mothers were themselves originally babies, it seems
that no universe will have an Arrow. Let us explain the argument
and see where it directs us.

\subsubsection*{{\textsf{3.1. The Universe as a Branch System}}}

When we observe a system in an unusually low entropy state, we
invariably find that it owes that low entropy to a process
involving a larger system. The \textit{overall} entropy of that
larger system need not be small; all that is needed is that (a)
the part that goes into making the smaller system should have low
entropy, and (b) the process of ``splitting" should not itself
greatly increase the entropy of that part. For example, if I wish
to cool my vodka, I go to the refrigerator (by no means a system
with low total entropy, by the standards of the glass of vodka)
and split off a low-entropy sub-system, to wit, a block of ice. I
convey this sub-system to my vodka, but I must take care to do
this without destroying its icy state. In principle one could
extract ice from a refrigerator using dynamite, but this is
unlikely to produce satisfactory results; the splitting of the
system \textit{alone} does not yield a refreshing low-entropy
drink: one needs to do it in the right way.

A low-entropy system obtained in this manner is said to be a
\textit{branch system} of the larger system. The existence of
branch systems is what allows the enormous complexity of our
world, despite the relentless increase of overall entropy demanded
by the second law of thermodynamics.

It is natural to guess that our own Universe may be a branch
system of some larger system, and that this may account for the
low initial entropy of the world we observe. Of course this does
not in itself solve the problem of the Arrow, but we may hope to
shift that problem to a venue where it may be more easily
addressed.

A good example of this way of thinking is provided by the theory
of Carroll and Chen \cite{kn:carroll} discussed earlier. Recall
that the idea here is that the observed Universe, among countless
others, splits off as a baby universe from some much larger and
older universe. That older world may have a large overall entropy,
but its entropy \textit{density} could be small; \emph{if} the
baby nucleates over a small region, and \textit{if} it is born in
the right way, it will begin with only a small amount of entropy,
in the manner typical of a branch system.

Now there are three fundamental questions which have to be
answered before this ``baby universe as a branch system" idea can
be said to work. First, of course, we have to ask why the larger
universe should have a low (\emph{geometric}) entropy density;
secondly, we have to determine whether, and in what precise sense,
baby universe nucleation does indeed take place in a small region
of spacetime; and finally, we must convince ourselves that the
splitting off of the baby does not increase its entropy to an
unacceptable degree.

\subsubsection*{{\textsf{3.2. The Importance of Being Baldest}}}

In the case of the Carroll-Chen theory, we can try to answer the
first question as follows. We try to use the idea, familiar from
discussions of Inflation, that an accelerated expansion of space
(due, for example, to the presence of positive vacuum energy) will
in some sense make the spatial sections more smooth. The intuitive
picture here is that space is like a rubber sheet; as it is
stretched, it becomes smoother. Presumably, if one is willing (and
able) to wait sufficiently long, this process will produce spatial
sections which are as smooth as we know the early spatial sections
of our Universe to have been. The end result of this kind of
smoothing is usually called \textit{cosmic baldness}.

Now in fact the process of cosmic balding is not as simple as the
``rubber sheet" analogy might suggest; see for example
\cite{kn:uggla} for a taste of the rather formidable
technicalities. It is \emph{not} true, in particular, that an
accelerating spacetime must globally come to resemble de Sitter
spacetime. The balding process \emph{only} takes place within the
cosmological event horizon of an individual inertial observer;
roughly speaking this is because of the exponential decay of the
proportion of a spatial slice accessible to a \emph{single}
observer. (See \cite{kn:gibsol} for a clear recent discussion of
this crucial point.)

In fact, this more subtle interpretation of cosmic baldness is
exactly what is needed for the Carroll-Chen theory to work. For
what it implies is that the mother universe, even if it
accelerates, never becomes isotropic at the global level: that is,
its ``total geometric entropy" is large. However, each local
inertial observer does see a low geometric entropy in his
immediate neighbourhood; the geometric entropy density is small.
Thus we see that cosmic baldness, properly interpreted, allows us
to answer the objection raised earlier against the Carroll-Chen
theory (that it did not appear to give rise to low
\emph{geometric} entropy in the baby universe).

On the other hand, one must be careful regarding the meaning of
``observer" here: in cosmology, an ``observer" can be an entire
galaxy or even a group of galaxies. Even the local smoothing-out
effect of the accelerated expansion only operates on larger length
scales than this, and, in fact, the observed acceleration in our
Universe is \textit{not} capable of, for example, tearing the
Milky Way apart. Actually, even the Local Group of galaxies can
resist the repulsion for an indefinite period \cite{kn:loeb}. Thus
the geometry of space within the Local Group will \emph{never} be
smoothed out by the accelerated expansion taking place on very
large scales, as long as the galactic matter remains within the
current confines of the Local Group.

Carroll and Chen, however, propose another way to scatter the
Local Group to the winds. They ask us to \textit{wait} until
essentially all of the matter in these galaxies has fallen into
black holes, and then to \textit{wait} for these black holes to
evaporate. Presumably the resulting radiation will cease to be
bound. In effect, this process contracts the ``observer" from the
size of the Local Group to smaller and smaller scales.
\textit{Eventually}, one might hope, the ``observer" will become
so small that the smoothing effect of cosmic acceleration can
operate over the tiny length scales on which baby universes
nucleate. After a further period of \textit{waiting}, this might
cause such a patch to become smooth on scales measured by
Penrose's number $P$. A baby which is born at this point has at
least some hope of being born with a geometric entropy as low as
that of the early stages of our Universe.

We have already expressed our reservations about whether one
should really be prepared, particularly if one is a string
theorist, to do this truly stupendous amount of waiting, and we
need not rehearse that discussion here. The point to be stressed
is that this is a very delicate construction, which could easily
malfunction \textit{even if} one is blessed with Boltzmannian
patience.

Nevertheless we do seem to be making some progress, so let us
proceed to the second question, concerning the extent of the
region over which baby universe nucleation takes place. Carroll
and Chen propose to rely on the mechanism of baby universe
nucleation proposed by Edward Farhi and Alan Guth \cite{kn:fargu}
(see also \cite{kn:aguirrejohn}; note that this mechanism differs
radically from the one usually employed in discussions of the
Landscape \cite{kn:TASI}). This kind of baby universe is what one
needs if, like Carroll and Chen, one wants to fluctuate
\emph{upwards} to a higher inflationary equilibrium. Now the idea
that Farhi-Guth baby universes are possible in string theory has
in fact been strongly challenged \cite{kn:hubeny}, and this is a
major problem for the Carroll-Chen theory. But let us leave that
to one side for the moment, and note that Farhi-Guth baby
universes do in fact branch off from a small spatial region in the
mother universe. If we can accept the argument that the latter has
achieved an ultra-low geometric entropy density by this time, then
the baby will indeed begin with the right properties if it can
maintain its spatial regularity through the birth process. This
brings us to the third question.

\subsubsection*{{\textsf{3.3. It's Not That Easy Being Born}}}

Farhi and Guth showed that, in the absence of exotic matter (see
below), the umbilical cord joining mother and child is swiftly
severed by a singularity. What this means is that the spatial
sections of the future baby shrink down to literally \emph{zero}
size in a finite time. The presence of a singularity is of course
unpleasant, but, for our purposes, the shrinking to zero size
itself is the real problem.

To see why, recall that, locally, expanding a universe tends to make
its spatial sections more uniform ---$\,$ this was precisely how we
are proposing to make the mother universe so smooth, as seen by an
individual observer. Equally, however, \textit{shrinking} the
spatial sections tends to make them highly anisotropic. Any small
blemish becomes increasingly apparent as the sections shrink, and
this happens very quickly when they become sufficiently small. It
turns out (see for example \cite{kn:turok}) that the size of a
spatial section can be quantified by a certain function of time,
$a(t)$, and that, when the sections are small, the anisotropy grows
according to $a(t)^{-\,6}$; this of course diverges rapidly if
$a(t)$ is allowed to tend to zero. (It diverges more rapidly than
almost anything that might try to defeat it; the exceptions
\cite{kn:turok}\cite{kn:podolsky2} are not relevant to the
Landscape.) If the spatial sections really shrink down to zero size,
then even the slightest anisotropy, even one measured on a scale
comparable to the reciprocal of Penrose's number, will be magnified
to an arbitrary extent. Our careful smoothing of space has been in
vain.

This is not very surprising: the birth of a baby universe (of any
kind) should be pictured as a highly traumatic event; certainly,
if the reader has any reason to suspect that a baby universe has
nucleated in his immediate vicinity, then he is urged to keep his
distance, if not to take to his heels. But this is just a way of
saying that we would \textit{not} expect the birth of a baby
universe to be an event that would give rise to a low-entropy
situation, or to preserve any low-entropy systems which pre-date
the birth. In other words, one simply expects that the birth of a
baby universe respects the second law of thermodynamics,
increasing the entropy as it proceeds. But, once again, ``entropy"
here means \emph{geometric} entropy; so we must expect that the
birth tends to make the relevant region less isotropic.

It has been argued \cite{kn:fischler} that quantum-mechanical
effects allow the singularity in the Farhi-Guth ``wormhole" to be
evaded; see \cite{kn:vacha} for a discussion of the merits of
this. In effect, the quantum fields behave like a kind of exotic
matter, ``exotic" in the sense that it violates the \textit{Null
Energy Condition}, the requirement that the sum of the energy
density and the pressure of a fluid should never be negative; see
\cite{kn:baby} for references.

We can picture this situation by imagining that the exotic matter
intervenes to prevent the spatial sections of the baby universe
from shrinking to \emph{exactly} zero size. We must still, in
accordance with the second law, expect a certain amount of
anisotropy to develop; but now at least we have some hope of
\emph{bounding} the increase. A detailed theory would have to show
that the increase in the anisotropy is sufficiently small as to
allow Inflation to start in the baby. The general point, however,
is that we can be sure that there \emph{will be} an increase. Thus
the initial smoothness of the mother universe must have been
\textit{even more extreme} than that of our own early Universe.

Many of these comments apply well beyond the particular scenario
posited by Carroll and Chen; they apply to any theory in which our
Universe appears as a branch system, splitting off as a baby
universe in a larger mother universe. The key point is that, in
any such theory, we must find some effective way of forcing the
mother universe to be or become extremely isotropic. In the case
of a baby connected to the mother by a wormhole, it suffices to do
this in the immediate vicinity of the place where the baby
nucleates; otherwise, the requirement has to be imposed on a much
larger region of the mother universe. In either case, the mother
needs to be even more isotropic than the baby. (Here and
henceforth, ``isotropic" means ``locally isotropic at the relevant
points".)

Since the Carroll-Chen theory is in any case committed to an
extreme version of the ``just wait" philosophy, the necessary
level of maternal isotropy can be achieved by waiting for cosmic
baldness to do its duty. (We assume that the entrance to the
wormhole remains small, so that its environs can always be
described by a single observer.) Thus, putting all of the many
pieces together, and assuming that they all work, we do now have a
theory of the Arrow. In summary: we wait for a Farhi-Guth baby
universe to nucleate in some small region of a mother universe
which does not have particularly uniform spatial sections but
which is accelerating. While we wait, cosmic baldness ensures that
the \emph{local} region becomes smoother, while collapse to black
holes and their subsequent evaporation effectively re-defines the
meaning of ``local", so that, when the baby does nucleate, it does
so in a region which is extremely smooth; so smooth that, even
when the anisotropy grows as the baby is born, the spatial
sections are still as smooth as Penrose demands. The baby universe
then has an Arrow arising from this extreme initial smoothness.

This is an ingenious and charming story, but we hope that we can
be excused for not believing in it. First, the colossal time
scales involved are likely to lead to very serious problems even
\emph{without} going into anthropic questions; when we do go into
them, extremely long time scales can only make the situation worse
(see for example \cite{kn:page}). Such long time scales do not
seem natural in string theory. Second, the mechanism whereby the
definition of ``local" is tremendously contracted, from
cosmological scales down to the size of a typical region in which
a wormhole can nucleate, does not seem to be sufficiently
reliable. Finally, it is very doubtful that Farhi-Guth baby
universes are compatible with string theory \cite{kn:hubeny}.

Nevertheless, the Carroll-Chen model does at least address all of
the problems we have identified, so it has much to teach us when
we consider other proposed ways of building universes. In
particular, of course, we are interested in building universes in
the string Landscape. Here, too, one uses baby universes, but of a
very different kind to the ones discussed above: the ones proposed
by Sidney Coleman and Frank De Luccia \cite{kn:deluccia}. This
kind of baby \emph{reduces} the value of the vacuum energy. It
nucleates in a small region but then expands very rapidly, for an
indefinite period, into the mother spacetime, instead of
disappearing into a wormhole. Its fate therefore cannot be
understood in terms of the observations of a single
observer\footnote{Technically: because the boundary of the baby
expands at a rate asymptotically approaching the speed of light,
the baby occupies a region of the relevant conformal diagram which
cannot be contained in the cosmological event horizon of a single
inertial observer.}, and so cosmic baldness is of \emph{no help}
to this kind of baby; the baby is no longer protected from outside
influences, a fact which has recently attracted some attention
\cite{kn:gagv}\cite{kn:aguijohsho}. Furthermore, the spatial
sections of the baby are \emph{infinitely large}; leaving aside
perturbations, each spatial section is a copy of the
three-dimensional space of constant negative curvature (as we
shall explain in more detail later). They therefore present a
large ``target" for outside influences, and in fact every part of
the earliest spatial sections is causally related to events deep
in the mother universe.

It is therefore clear that, as in the case of Farhi-Guth baby
universes, the second law of thermodynamics will apply here: if
the mother universe is not smooth on large scales, nor will the
baby be smooth; in fact it will be less so\footnote{As in the
Farhi-Guth case, the growth of anisotropies has to be controlled
by means of violations of the Null Energy Condition; see
\cite{kn:baby}.}. The situation here is in fact even worse,
because we cannot appeal to cosmic baldness. This means that the
\emph{total} entropy of the mother universe has to be extremely
low if that of the baby is to be low: the second law of
thermodynamics can no longer be circumvented, as it was in the
Carroll-Chen theory, by contrasting total entropy with entropy
\emph{density}.

To be brief: baby universes are an important feature of the
Landscape, but in themselves they do not help us to understand the
Arrow of Time. In fact, they merely recapitulate the problem on a
higher level.

\subsubsection*{{\textsf{3.4. Haunted by the Past}}}
A baby in the Landscape, then, will have an Arrow of Time only if
its mother does: it has to \emph{inherit} an Arrow. Of course, it
is all too clear that we have not solved the problem of the Arrow
in this way. Actually, we may have made it worse, because the
mother has to be at least as smooth as the baby. Furthermore, the
mother itself may well be the baby of some still larger
``grandmother universe", which was as smooth or smoother still;
and so on.

In our view, we have arrived at this impasse because we have tried
to avoid the conclusion of our discussion in Section 2.2. In fact,
we have tried to use the baby universe concept to avoid speaking
of the \emph{beginning} of time in any real sense. This has not
worked. Intuitively, anything which \emph{has a past} will be
influenced by that past; it cannot be as smooth as Penrose's
calculation demands, unless its past was at least as smooth. The
most straightforward conclusion is that Penrose's calculation
really means that our Universe ultimately arose from some system
which simply \emph{had no past}. In the context of baby universes,
this means that we must trace the origin of the Arrow back through
the mother universe, the grandmother, and so on, until we reach
the original female ancestor. Let us call this original universe
``Eve". By definition, then, Eve had no past whatever. This made
it possible, in some way to be explained, for Eve to begin with an
Arrow; all subsequent generations have an Arrow \emph{only}
because they inherited one from Eve.

Our task, then, is to explain how Eve acquired her Arrow.

\section*{\large{\textsf{4. All About Eve}}}
Eve had no past, but nevertheless was subject to some kind of
physical law. How can that be? Can a universe really have no past?

Ideas of this sort first arose when Alexander Vilenkin
\cite{kn:vilenkin} made the daring suggestion that the Universe
was created literally from \emph{nothing}. (The celebrated ``no
boundary" theory of Hartle and Hawking \cite{kn:hartle} can also
be interpreted in a similar way.) It is, to put it mildly,
difficult to think about ``the physics of nothing", but string
theory suggests a way of doing so. It is generally felt
\cite{kn:polchinski} that, in some sense which today remains
rather vague, time itself is \emph{emergent} in string theory. Of
course, much remains to be done to render this idea more precise,
but the rough idea is that if one probes sufficiently far back in
time, one will reach a region beyond which time simply ceases to
be meaningful. Evidently, that region cannot have a past.
According to our discussion, the boundary, which we necessarily
picture as a spacelike hypersurface, is where we must seek the
real origin of the Arrow.

The stringy ``pre-emergence" state is not an \emph{alternative} to
Vilenkin's idea, so much as a way of giving a deeper analysis of
``nothingness". It certainly does seem problematic to speak of the
Universe ``existing" in the absence of time; see \cite{kn:vaas}
for philosophical background on such questions. We shall therefore
continue to speak of the Universe being \emph{created} or
\emph{destroyed} along the surfaces which are boundaries between
regions in which time is well-defined and regions in which it is
not. Let us try, then, to find a stringy account of creation.

\subsubsection*{{\textsf{4.1. The Minimal Torus of Creation}}}

What would ``spacetime" be like if time did not exist? There is an
apparently naive yet ultimately profound answer to this question;
it runs as follows. Take Minkowski spacetime, the usual spacetime
of Special Relativity. Distances in Minkowski spacetime are
measured using the Minkowski metric, given by
\begin{equation}\label{eq:F}
g^M_{-\,+\,+\,+} \;=\;-\,dt^2\;+\;dx^2\;+\;dy^2\;+\;dz^2\,,
\end{equation}
where t denotes time, the other coordinates are the familiar
Cartesian ones for the spatial dimensions, and the pattern of
signs, $-\,+\,+\,+$, is called the \emph{signature}. Time is the
dimension corresponding to the aberrant sign in this formula.
``Spacetime without time" would be the ``$+\,+\,+\,+$" version of
this formula,
\begin{equation}\label{eq:G}
g^M_{+\,+\,+\,+} \;=\;+\,dt^2\;+\;dx^2\;+\;dy^2\;+\;dz^2\,.
\end{equation}
This is described \cite{kn:hartle} as the \emph{Euclidean} version
of the original spacetime metric. (The terminology is unfortunate:
it means nothing more than that the metric has $+\,+\,+\,+$
signature; in particular, it certainly does \emph{not} mean that
the metric is flat.) Naively, then, we expect time to emerge from
a stringy state with Euclidean signature. In terms of this
oversimplified example, this means that time emerges at the point
at which $g^M_{+\,+\,+\,+}$ is suddenly\footnote{We would argue
that it is analytic that time cannot ``emerge gradually". But see
\cite{kn:mars}.} replaced by $g^M_{-\,+\,+\,+}$. Notice that each
formula can be obtained from the other by replacing $t$ by $it$,
where $i$ is the usual imaginary unit, $\sqrt{-\,1}$. One says
that this coordinate has been \emph{complexified}.

Thus we obtain a concrete picture of emergent time. This is a
useful idea, however, only if we can say something about the
``pre-temporal", Euclidean region.

Vilenkin and Hartle and Hawking made proposals as to the structure
of the Euclidean region; more recently, a radically different
suggestion, based on string theory, was made by Hirosi Ooguri,
Cumrun Vafa, and Erik Verlinde \cite{kn:ooguri}. They begin far
from cosmology, by studying a kind of stringy black hole, leading
to a two-dimensional Euclidean metric of the form
\begin{equation}\label{eq:H}
g^{OVV}_{++}\;=\;K^2\,e^{(2\,\rho/L)}\,d\tau^2\;+\;d\rho^2.
\end{equation}
Here one should think of $\rho$ as a coordinate which runs along a
line, but of $\tau$ as the angular coordinate giving position on a
circle, of radius related to the fixed number $K$. (The constant
$L$ measures the curvature of this space.) Now the Euclidean
``time" here
---$\,$ that is, the dimension which would normally be complexified
---$\,$ is $\tau$. (It is usual for black hole Euclidean ``time" to be
compactified in this manner; the periodicity of ``time" is
ultimately related to the thermodynamics of the black hole.) But
Ooguri et al. noticed that this metric looks very much like a
\emph{cosmological} metric: if we take $\rho$ to be Euclidean
``time", then the geometry is like a Euclidean version of an
exponentially expanding two-dimensional cosmos with \emph{finite}
(circular) ``spatial" sections. The fact that a single Euclidean
space can have \emph{two} spacetime interpretations opens the way
to a profound application of stringy black hole theory to
cosmology.

The four-dimensional version of this geometry is just
\begin{equation}\label{eq:I}
g^{OVV}_{++++} \;=\; d\rho^2\;
+\;K^2\,e^{(2\,\rho/L)}\,(\,d\theta_1^2 \;+\; d\theta_2^2 \;+\;
d\theta_3^2).
\end{equation}
Here $\theta_{1,\,2,\,3}$ are angular coordinates on three
separate circles. Such a product of circles is called a
three-dimensional \emph{torus}. It may be pictured as the interior
of a box with non-trivial topology; as one tries to exit through a
side, one instantly re-appears inside the box, entering it from
the opposite side; one says that the opposite sides have been
\emph{topologically identified}. Thus, the cosmological version of
this space will have spatial sections each of which is a copy of a
torus.

The torus has a property which turns out to be crucial here: it is
finite (compact) yet \emph{flat}. That is, inside the box, all of
the usual laws of flat-space geometry obtain; the space differs
from ordinary flat space \emph{only} through its topology. That is
why the metric $g^{OVV}_{++++}$ contains two independent length
scales: $L$, which fixes the overall curvature in four dimensions,
cannot control the sizes of the ``spatial" sections (determined by
$K$), precisely because these sections have no curvature of their
own. Since the length scale of the torus is independent of $L$ for
this particular (exactly flat) spatial geometry, it must be
independent of $L$ even if we begin to distort the torus away from
exact flatness. (It should be pointed out here that string theory
is defined on higher-dimensional spacetimes, and that the
``hidden" dimensions are always taken to be compact, and often to
be flat. Thus, the compactness of the ``visible" dimensions is
particularly natural in string cosmology, a fact strongly
emphasised by Ooguri et al. \cite{kn:ooguri}.)

Now let us follow Ooguri et al. and take this metric as the
Euclidean version of a spacetime metric. That is, we truncate the
space along the three-dimensional surface $t = 0$, where time
emerges; notice that this surface has the topology of a torus. The
spacetime on the other side has a metric which is obtained by
complexifying both $\rho$ and $L$ (so that $e^{2\,\rho/L}$ remains
real). This works \emph{only} because $K$ is an independent
parameter; if the ``spatial" sections had themselves been curved,
this curvature would have been determined by $L$, with disastrous
consequences when $L$ is complexified. For these simple yet deep
reasons, we maintain \cite{kn:OVV} that the Ooguri-Vafa-Verlinde
account of the ``pre-temporal" state \emph{requires} time to emerge
along a torus.

It is surely no coincidence that spaces of toral
topology\footnote{There are actually a few other compact flat
three-dimensional manifolds apart from the three-torus, but all of
them can be obtained from the torus by means of yet further
topological identifications. Henceforth, ``torus" means any member
of this finite and fully-understood family of spaces.} actually
play a very basic role in string theory. String theory has a very
unusual symmetry, known as T-duality \cite{kn:polchbook}. In
general terms, this symmetry implies that string theory defined on
a torus is insensitive to a replacement
$K\;\rightarrow\;L^2_{string}/K$, where $K$ is the toral radius
parameter we have been discussing, and where $L_{string}$ is a
fundamental parameter of string theory, the \emph{string length
scale}. ``Insensitive" here means that the result of this
transformation is the same physics described in a different way.
This is often interpreted to mean that, in string theory, it does
not make sense for a torus to have a ``radius" smaller than about
one string length. (This slightly vague idea is made explicit in
one particularly interesting approach to string cosmology, the
``string gas cosmology"; see \cite{kn:brandenberger}.) This
distinguished torus of \emph{minimal} size is the obvious
candidate for the surface along which time emerged. (In fact,
earlier work on ``creation from nothing"
\cite{kn:vilenkin}\cite{kn:hartle} always did assume that the
Universe was created along a minimal spacelike surface; here we
are saying that this assumption is particularly natural and
meaningful in string theory defined on a torus.)

Notice that the existence of a minimal initial spacelike surface
automatically solves the problem of having an initial Big
Bang-style singularity; that is, the theory implicitly violates
one or another assumption of the \emph{Singularity Theorems.} In
fact a toral cosmology can only be non-singular by violating the
Null Energy Condition; but we are already committed to this in any
case if we intend to make use of baby universes. (Exotic matter
which violates the Null Energy Condition will violate any of the
conditions assumed by the Singularity Theorems.) The original
space considered by Ooguri et al. (with metric given by equation
(\ref{eq:I})) does \emph{not} contain any minimal surface; the
exotic matter removes the singularity and replaces it by a
geometry which does have such a surface.

It is implicit in this argument that it is possible for the
Universe to come into existence with a size given approximately by
the string scale. That seems very reasonable, but we should point
out that the string scale is normally thought to be substantially
larger than the \emph{Planck scale}, the scale at which it is
usually surmised that \emph{quantum gravity} effects come to
dominate. This suggests that quantum gravity does not in itself
fully explain the Arrow. (It does play a role in the theory to be
put forward later, but only through small, perturbative effects.)
Once again, the fact that the toral scale $K$ is completely
independent of the spacetime curvature scale $L$ is of basic
importance here.

On the other hand, the string scale may be too \emph{short} to be
the scale at which Inflation starts. Happily, in the cases of
interest to us, the toral topology \emph{itself} allows this
problem to be solved, in the following manner \cite{kn:arrow}.
Essentially what happens is that the toral topology initially
prevents cosmological horizons from forming, so that the entire
universe can be seen by a single observer. (Recall that a
cosmological horizon forms when objects escape beyond the
observer's view; but it is difficult to escape from a torus, since
the escapee tends to find itself back where it started.) Cosmic
baldness then applies to the entire universe, so that, if we can
achieve smoothness \emph{initially}, this smoothness will be
maintained by that effect. Eventually, however, a cosmic horizon
does form, and Inflation in the conventional sense can begin; but,
by that time, the universe has expanded to the appropriate size,
much larger than the string scale. Thus, toral topology can
reconcile the Inflation scale with the string scale. (See
\cite{kn:lindetypical} for the use of tori to ``delay" Inflation,
though the approach there is different.)

Throughout this discussion, we have been assuming, for simplicity,
that the torus is endowed with its \emph{exactly flat}, and
therefore perfectly locally isotropic, geometry. In discussing the
Arrow in this context, we cannot of course begin in this way.
Fortunately, everything we have said here can be formulated for a
completely \emph{arbitrary} geometry on the initial torus, fixing
only its topology. With such an arbitrary geometry, it no longer
makes sense to speak of the ``radius" of the torus, but it still
has a well-defined (finite) volume; this fixes a definite length
scale, which we continue to denote by $K$. The overall spacetime
in the immediate vicinity of the initial spacelike surface will
also have a characteristic length scale, independent of $K$, which
again we continue to denote by $L$. The concept of a \emph{minimal
surface} in a general Riemannian manifold is a classical one,
defined as follows \cite{kn:kobayashi}. Consider a family of
finite (compact) subspaces embedded in a manifold. The subspaces
have metrics induced on them, and, as one moves from one subspace
to another, this metric will usually change, and so will the
volume of the subspace. The rate of change of the metric is
measured by an object usually called (in physics) the
\emph{extrinsic curvature}; it can be regarded as a tensor $K^a_b$
defined in each subspace. A subspace is said to be \emph{minimal}
if it minimizes the volume for variations with fixed boundaries,
and this requires the trace of this tensor to vanish: we have:
\begin{equation}\label{eq:J}
K^a_a\;=\;0,
\end{equation}
with an implied summation on $a$.

The concept of a minimal surface can be generalized to the
situation we are considering here, resulting in the same equation.
We claim that time emerges along a spatial slice of minimal
volume, a claim which we can now formulate in terms of this
equation. (In this discussion we have used \emph{Riemannian}
geometry, meaning that we have been examining the boundary surface
``from the Euclidean side", but the surface will also be minimal
from the spacetime side, in the obvious sense.)

On the Euclidean side, the geometry we are discussing can be
described concretely as follows. The vectors perpendicular to the
boundary minimal surface allow us to define a coordinate $\rho$
such that the metric, at least locally, takes the form
\begin{equation}\label{eq:K}
g(F,\,h_{a\,b},\,K,\,L)
\;=\;d\rho^2\;+\;K^2\,h(\rho/L,\,\theta_c)_{a\,b}\,d\theta_a\,d\theta_b\,,
\end{equation}
where the $\theta_{1,2,3}$ are the angular coordinates on the
torus, and $h(\rho/L,\,\theta_c)_{a\,b}$ is the metric on the
torus labelled by $\rho$; it is completely arbitrary, apart from
the condition that its dependence on $\rho$ is subject to the
demand that the boundary must be minimal. The corresponding
spacetime metric on the other side of the boundary is obtained
simply by complexifying $\rho$ and $L$ (but not $K$).

In this section, we have argued that the Arrow is related to the
\emph{emergence} of time itself. The argument is simple: that
which has no past cannot be distorted by any ``prior" conditions.
Guided by the work of Ooguri et al. on the string-theoretic view
of the ``pre-temporal" state, we have been led to a concrete
picture of the earliest era of the universe. It would certainly be
of great interest to pursue the programme of Ooguri et al., in
order to further elucidate the nature of the Euclidean region.
(Notice for example that the complexification of $L$ means that,
if the spacetime is predominantly positively curved, as it will be
if it inflates, then the Euclidean region must be
\emph{negatively} curved; indeed, Ooguri et al. use a space which
is locally identical to hyperbolic space.) However, that is not
needed for our purposes.

We have seen that string theory may contain a ritual for
exorcising the ghosts of the past. The penalty we pay for this
exorcism is however a heavy one: whatever the principle
responsible for Eve's Arrow may be, it cannot, for obvious
reasons, act causally. In the next section we argue that general
relativity contains certain features which, in a sense, \emph{are}
acausal.

\subsubsection*{{\textsf{4.2. The Initial Value Problem for Gravity}}}
We have repeatedly referred to Newton's epochal separation of
\emph{dynamical laws} from \emph{initial conditions}. For Newton
this was a natural thing to do, because the initial positions and
momenta of a collection of classical particles can be prescribed
completely arbitrarily. Gravity, however, is different.

At first sight it seems paradoxical to speak of ``time evolution"
for gravity, since, if one does not \emph{already} know the
spacetime structure, one does not know how time itself behaves.
This apparent paradox is resolved in the following extraordinary
way. One begins with a three-dimensional manifold $\Sigma$, on
which is given a Riemannian (that is, signature +,+,+) metric
$h_{ab}$, a symmetric three-dimensional tensor $K_{ab}$, a
function $\varrho$ and a three-dimensional vector field $J^{a}$,
together with data for the ``matter fields". At this point,
$K_{ab}$, $\varrho$, and $J^{a}$ ``have no meaning"; one simply
proceeds to solve the Einstein field equations in a formal way, so
that $\Sigma$ is embedded in the resulting spacetime. One
\emph{then} finds that these fields ``turn out to have been",
respectively, the extrinsic curvature of the ``initial" surface
$\Sigma$, the value of the energy density of the matter fields on
$\Sigma$ as measured by observers whose worldlines are
perpendicular to $\Sigma$, and the projection onto $\Sigma$ of the
four-dimensional \emph{energy-momentum flux vector} seen by these
observers. Thus these objects acquire their meanings
\emph{retrospectively.}

All of this is strange and suggestive enough. But what is still
more interesting is the following. Suppose that we try to
prescribe the metric on $\Sigma$, the tensor $K_{ab}$, and so on,
completely independently of each other. Then, generically, one
will find that the Einstein field equations simply \emph{do not
have a solution.} Solutions exist only if the initial data are
related to each other by means of a complicated set of equations,
the \emph{initial value constraints}:
\begin{eqnarray}\label{eq:L}
D^a\,(K_{ab}\;-\;K^c_c\,h_{ab}) & = & -\,8\pi J_b \ \nonumber \\
              R(h)\;+\;(K^a_a)^2\;-\;K_{ab}\,K^{ab} & = &
              16\pi\varrho.
\end{eqnarray}
Here $D^a$ is the covariant derivative operator in $\Sigma$, there
is an implied summation on the indices when they are repeated, and
$R(h)$ is the \emph{scalar curvature} defined by $h_{ab}$. (See
\cite{kn:waldbook}, Chapter 10. We can regard $D^a$ as an operator
which measures spatial rates of change \emph{within} $\Sigma$; we
shall discuss the meaning of $R(h)$ in more detail later.) One
must not think of these relations as having anything to do with
the fields interacting with each other; they are strictly acausal;
they are imposed on the initial data by the presumed existence of
the ``subsequent" spacetime as a structure compatible with the
Einstein field equations.

Now the following idea suggests itself. The function $\varrho$ has
(retrospectively) a direct physical significance, and it is
reasonable to suppose that some physical principle will restrict
it. Later, for example, we shall suggest that $\varrho$ should
never be negative along a surface $\Sigma$ where a universe is
being created. The initial value constraints then impose
conditions on the geometry of $\Sigma$. One might even regard
these conditions as \emph{laws of nature} of precisely the kind
requested in Section 2.2. In fact, however, this point of view has
never been taken. Why?

The reason can be understood by considering a special case, the
beginning of time as it appears in the Hartle-Hawking theory
\cite{kn:hartle}. Here time begins (or emerges) along a spatial
section with vanishing extrinsic curvature and with the topology
of a three-dimensional sphere. The second equation in (\ref{eq:L})
is then simply
\begin{eqnarray}\label{eq:M}
R(h)\;& = &\;
              16\pi\varrho.
\end{eqnarray}
Thus the scalar curvature and the initial energy density cannot be
prescribed independently. But does this really represent much of a
restriction? Suppose for example that we are given $\varrho$ as an
arbitrary non-negative function on the three-dimensional sphere.
Can we find a metric such that (\ref{eq:M}) is satisfied? The
answer is that this is, indeed, \emph{always} possible: this is a
consequence of a profound theorem due to Jerry Kazdan and Frank
Warner \cite{kn:kazdan}; see \cite{kn:arrow} for a discussion of
this theorem. If $\varrho$ is an extremely complicated, irregular
function, then the geometry of $\Sigma$ will likewise be extremely
irregular. In short, under normal circumstances, the geometry of
the topological three-sphere is constrained so weakly by the
initial value constraints that the latter cannot be considered
worthy of the title of ``laws of nature".

So it is, at least, for universes created with spherical topology.
One of our main claims in this work is that it is otherwise with
the torus. Before explaining this, let us be more precise as to
the physical conditions we are going to assume.

\subsubsection*{{\textsf{4.3. Creation vs Destruction}}}
In order to explain the origin of the Arrow, we must of course
make some physical assumptions. These should either follow from
string theory, or at least be so plausible (or ``mild") that one
can hold out hope that they \emph{will} follow from string theory
(when, perhaps, the emergence of time is better understood).

In general relativity, each observer defines at each point a
four-dimensional \emph{energy-momentum flux vector} $T^{\mu}$.
This specifies the flow of energy in various spatial directions,
as seen by this observer. Also, the observer's worldline has a
tangent vector $V^{\mu}$ at each of its points. In general these
two vectors need not be parallel, but one can specify whether they
point in the same general direction in spacetime by computing the
projection of $T^{\mu}$ onto $V^{\mu}$; this projection, if it is
not zero, is either parallel to $V^{\mu}$ or anti-parallel to it
(that is, it points in exactly the opposite direction.)

Now consider the spacelike hypersurface $\Sigma$ constituting part
of the boundary of spacetime, and let $N^{\mu}$ be the field of
unit vectors, perpendicular to $\Sigma$ at each point, and
pointing \emph{inwards} from the boundary. (The inwards/outwards
distinction can be given a rigorous mathematical definition; see
\cite{kn:arrow}.) We can think of $N^{\mu}$ as defining tangent
vectors to the worldlines of a family of observers. Suppose now
that we claim that this particular $\Sigma$ represents the
\emph{creation} of the Universe. Whatever this means precisely, we
can formulate it mathematically as follows. If $T^{\mu}$ is the
energy-momentum flux vector field associated with $N^{\mu}$, then
$T^{\mu}$, evaluated on $\Sigma$, \emph{never points outwards};
that is, its projection on $N^{\mu}$ is never anti-parallel to
$N^{\mu}$. On the other hand, if $\Sigma$ is a boundary along
which the Universe is being destroyed, then we shall take this to
mean that $T^{\mu}$, again evaluated on $\Sigma$, does point
\emph{outwards}. The intuitive picture here
---$\,$ not to be taken too literally
---$\,$ is that while the spacelike components of $T^{\mu}$ measure
flows of energy within $\Sigma$, its timelike component tells us
whether energy is ``coming into the world" (in a direction
perpendicular to $\Sigma$) or taking leave of it.

These statements are the \emph{only} input we shall need regarding
the energy content of the universe. In order to \emph{prove} them,
one would need to have a detailed physical theory of ``creation"
and ``destruction", which in turn would demand an understanding of
the ``emergence of time" far beyond what we can claim to have. At
this point they are just simple postulates about the correct
mathematical representation of ``creation" and ``destruction".
They are motivated by the Landscape, in the following sense.

Recall that, in the Landscape, the nucleation of a Coleman-De
Luccia baby universe has the effect of \emph{lowering} the value
of the vacuum energy. Certain unfortunate babies will ``overshoot"
the desired value of the cosmological constant, so that they have
a \emph{negative} vacuum energy. These will expand at first, but
eventually they will begin to contract; classically, they will
terminate in a singularity. (A vacuum which contains nothing but
negative vacuum energy is never singular; this is the famous
``anti-de Sitter space" beloved of string theorists. If, however,
such a spacetime also contains a small amount of ``normal" matter,
then the collapse will not be so innocuous.) We shall assume that,
instead of a singularity, a string-theoretic treatment will lead
to a minimal spacelike surface, along which time ``submerges",
that is, it ceases to be meaningful. Now despite their name, the
``initial" value constraints apply in this case; substituting the
minimality condition (equation (\ref{eq:J})) into the second
member of (\ref{eq:L}), we obtain
\begin{eqnarray}\label{eq:LL}
16\pi\varrho\;=\;R(h)\;-\;K_{ab}\,K^{ab}.
\end{eqnarray}
Now recall from section 3.3 that the spatial sections of a
Coleman-De Luccia baby universe are, if perturbations are ignored,
\emph{negatively} curved. Therefore, we see from this equation
that $\varrho$ has to be negative in this case, since the term
involving the extrinsic curvature can be expressed, at each point,
as a sum of squares. But $\varrho$ is just the time-component of
$T^{\mu}$ as seen by observers with inward-pointing tangent
vectors (see \cite{kn:waldbook}, Chapter 9); its being negative
implies that $T^{\mu}$ points outward. Hence we see that,
according to our postulates, spacetime is indeed being
\emph{destroyed} in this case; which is as it should be.

This discussion is for \emph{motivation}: it does not prove
anything. For, as always, we must not \emph{assume} that the
spatial sections of this universe will retain the geometry of an
exact space of constant negative curvature. If they did so, then
in fact $R(h)$ would become \emph{more} negative as this universe
shrinks. If they fail to do so, then the behaviour of $R(h)$ is
less clear; on the other hand, the term involving the extrinsic
curvature would then grow, so in either case we can expect
$\varrho$ to remain negative. Also, independently of this
equation, one can argue on general grounds that there must be a
strong \emph{negative} contribution to $\varrho$ (in addition to
the background vacuum energy, which of course is also
\emph{negative}) from whatever it is that prevents a singularity
here. Thus, it does seem that, in the Landscape, the destruction
of universes is associated with negative (total) energy densities.
Our postulate, based on this evidence, is that this is true in
general. Conversely, it is natural to suppose that the total
energy density is never negative when a universe is created. (Note
that the vacuum energy in the creation case is \emph{not}
negative, and that the topology and geometry of the spatial
surfaces are also different in that case.)

We stress the obvious: the conditions we are proposing here are
almost \emph{ridiculously weak}, especially when they are compared
with the extremely strong restrictions on the geometry of $\Sigma$
which we hope to extract from them. All they require is that the
vector $T^{\mu}$ should point in a certain \emph{general}
direction; this is no more specific than saying that (either)
Cambridge is ``northward" of Singapore. Furthermore, they require
this only on $\Sigma$, not in the bulk of the spacetime itself.
This ``weakness" is very important, because it means that our
arguments are correspondingly robust. If we had demanded physical
conditions involving \emph{equations}, then we would have to fear
that corrections to those equations might ruin the argument. But
all we are requesting here is that the time component of $T^{\mu}$
should satisfy a certain \emph{inequality} along $\Sigma$.
Obviously, inequalities are robust: if a certain number is
positive, for example, then so are all nearby numbers. We will
shortly see how this works in detail.

\subsubsection*{{\textsf{4.4. And Then A Miracle Happens: The Arrow Explained}}}
Now let us put all of the pieces together. We assume that the Eve
universe was \emph{created} along a \emph{minimal-volume}
spacelike surface $\Sigma$ with the topology of a
three-dimensional torus but with otherwise unknown geometry
(except that the overall length scale is given by
$K\;\approx\;L_{string}$). Why should we believe that the geometry
of this torus is not generic, that is, extremely irregular?

First, note that we agreed that ``being created" means that
$T^{\mu}$, the energy-momentum flux vector field defined by the
inwards-pointing normal field $N^{\mu}$, should never point
outwards. But, because $\varrho$ is the time-component of
$T^{\mu}$ as seen by these observers, this simply means that,
along $\Sigma$ only, we have
\begin{eqnarray}\label{eq:N}
\varrho\;& \geq &\; 0.
\end{eqnarray}
Inserting this and the minimality condition $K^a_a\;=\;0$ into the
second member of the initial value constraint equations
(\ref{eq:L}), we get
\begin{eqnarray}\label{eq:O}
R(h)\;=\;K_{ab}\,K^{ab}\;+\;
              16\pi\varrho \;\geq\;0;
\end{eqnarray}
recall that the term involving the extrinsic curvature is
essentially a sum of squares.

As expected, our weak physical assumptions have led to an
apparently feeble constraint on the metric on $\Sigma$. First, as
we have stressed, we have a mere inequality. Second, let us recall
the definition of the \emph{scalar curvature}. At each point, a
three- (or higher-) dimensional manifold can curve in many ways,
depending on the \emph{direction} in which the space is sliced.
Thus, even at a fixed point, such a space can be negatively curved
in one direction, and positively curved in another. The scalar
curvature at a point is simply a constant multiple of the
\emph{average} of all of these various curvatures. Now to say that
the average of a collection of numbers is not negative is to say
almost nothing: for example, all of them but one could be
negative, as long as the one exception is sufficiently large that
it outweighs all of the others combined when the total is taken.
Thus, we have no right to expect anything interesting from a
condition like (\ref{eq:O}); and, as we mentioned earlier, in the
case of \emph{spherical} topology one can actually prove that this
condition hardly constrains the geometry at all.

But now something extraordinary arises. One has the following
theorem, obtained by combining several extremely deep geometric
results of Richard Schoen, Shing-Tung Yau, Mikhail Gromov, Blaine
Lawson, and Jean-Pierre Bourguignon. (For technical details, and
full references, see \cite{kn:lawson} and \cite{kn:arrow}.)

\bigskip
\noindent \textsf{THEOREM (Schoen-Yau-Gromov-Lawson-Bourguignon)}:
Consider the set of \emph{all possible Riemannian metrics} on any
torus. In this set, the only metrics with everywhere non-negative
scalar curvature are those which are perfectly flat; that is,
their \emph{full curvature tensor} vanishes exactly.
\bigskip

\noindent The proof of this theorem ---$\,$ let us call it the
SYGLB theorem for brevity ---$\,$ is very abstruse; appropriately,
perhaps, it makes use of techniques originally drawn from physics
(particularly the concepts of \emph{Dirac operators} and
\emph{spin geometry}).

The SYGLB theorem is a genuinely astonishing result. First, the
theorem surveys \emph{all} possible geometries on the torus.
Second, and even more startling, it means that the \emph{average}
of the curvatures can only be non-negative if each and every
curvature vanishes exactly
---$\,$ the curvatures cannot ``average out" to a non-negative
value. Third, the result must somehow link the curvature tensor to
the specific topology of the torus, since no such result holds in
the case of spherical topology.

Suddenly we find ourselves in a different world from the case of
the sphere. For now the \emph{inequality} (\ref{eq:O}) implies the
following \emph{equations}.

First, and of course most importantly, (\ref{eq:O}) now implies
that the full curvature tensor of the ``initial" three-dimensional
surface $\Sigma$ is exactly equal to zero. The Eve universe had to
be created along a spacelike surface which was perfectly flat,
which means that it was exactly locally isotropic around each
point.

This is of course a major step forward, but we are not yet done.
Essentially, we have fixed the initial spatial metric, but we have
not yet fixed its ``time derivative", as we must do if we wish to
have a complete set of initial conditions. Furthermore, while it
may be ``natural" to assume that the initial state of the inflaton
shares the symmetries of the initial spatial section, ideally we
should \emph{prove} this. In principle, the geometry might be
perfectly regular while the inflaton field is highly irregular.

Again we refer to (\ref{eq:O}): having deduced that the curvature
is zero, the left side must vanish, and so therefore must the
right. But since the right side is a sum of terms which cannot be
negative, the only way they can add to zero is if all of the terms
vanish separately. Thus we find that
\begin{equation}\label{eq:P}
K_{a\,b}\;=\;0;
\end{equation}
that is, the entire extrinsic curvature matrix vanishes initially.
This is of course a far stronger condition than the mere vanishing
of its trace. The initial ``time derivative" of the spatial metric
is now fixed: it has to vanish. This implies that the initial
moment was a \emph{moment of time symmetry} (\cite{kn:waldbook},
Chapter 10). \emph{Classically}, this forces the time derivative
of the inflaton field to vanish initially, an important
contribution to putting the inflaton into its minimal-entropy
state. Of course, the geometry on the other side of $\Sigma$ is
not really the same as that on the spacetime side; but
\emph{classical} local fields would not ``know" this. Quantum
effects spilling over from the Euclidean domain may not be totally
negligible, however, so perhaps it would be more prudent to say
that this argument shows that the initial time derivative of the
inflaton must be \emph{very small}. This is acceptable, because
the inflaton need not be exactly constant for the approximate
relation (\ref{eq:E}) to hold. (In fact we \emph{want} the
inflaton to start (slowly) ``rolling".)

Next, we consider the \emph{first} equation in (\ref{eq:L}). Now
that we know that the extrinsic curvature vanishes everywhere on
$\Sigma$, so do its derivatives, so this equation simply requires
that
\begin{equation}\label{eq:Q}
J^a\;=\;0.
\end{equation}
Recall that $J^a$ is the projection into $\Sigma$ of the
energy-momentum flux vector. This implies that energy cannot flow
in any direction at any point of $\Sigma$, which just means that
the inflaton field is uniformly distributed over $\Sigma$: it does
share the uniformity of the underlying space, and so the spatial
derivatives of the inflaton field vanish, and the inflaton is
indeed initially in its low-entropy, potential-dominated state.

Finally, we now see from (\ref{eq:O}) that the total energy
density must vanish exactly everywhere on the initial surface.
This is a reminder that some kind of exotic effect, which violates
the Null Energy Condition, must act initially in order to
invalidate the Singularity Theorems; it must contribute a negative
energy density which exactly cancels the energy of the inflaton
initially (but not thereafter). A discussion of the nature of this
effect would take us too far afield here, but we may mention that
the \emph{Casimir effect} arises naturally in spaces with toral
topology; see \cite{kn:coule} for a general discussion of Casimir
effects in cosmology, and \cite{kn:nimah1}\cite{kn:nimah2} for
recent discussions from a string-theoretic point of view. The
negative energy density need not be precisely a Casimir energy,
but it must be something similar: like Casimir energy, it must
dilute very rapidly as the Universe expands, so that, after a
short period of expansion, its effect on inflationary physics is
negligible. (Lest the reader suspect a circular argument here,
note that the initial vanishing of the total \emph{energy density}
by no means implies the initial vanishing of the total
\emph{pressure}. Under reasonable physical conditions, the
``primordial pressure" will in fact be negative, and this will
launch the newborn Eve into an immediate phase of accelerated
expansion, even though the inflaton energy density has been
(momentarily) canceled. For speculations on the possible identity
of the ``Casimir-like" effect, see \cite{kn:arrow}.)

We can now claim to have fixed the initial conditions of the Eve
universe, in a manner which is so restrictive that, as requested
in section 2.2, it deserves the name of a ``law of nature". For
now the initial metric has been fixed (up to relatively
inconsequential issues such as the precise global shape of the
initial flat torus), and so has its initial ``time derivative"
(equation (\ref{eq:P})). The (classical version of the) law simply
states that the Eve universe was born along a surface which was
perfectly flat and embedded in such a way that its extrinsic
curvature was exactly zero. The law has been derived from the
string-motivated demands that the topology of this surface should
be toral (from the construction of Ooguri et al.
\cite{kn:ooguri}), that it should be minimal (from T-duality), and
that the energy-momentum flux vector on it should never point
outwards from spacetime.

This is the origin of the Arrow of Time: the initial geometry has
to be smooth, the inflaton has to begin in the potential-dominated
state. The Eve universe (eventually) inflates, and gives birth to
baby universes which inherit her Arrow. The rest is history.

We conclude this section by asking: is there anything analogous to
the SYGLB theorem for other compact three-dimensional manifolds?
This problem can be solved with the aid of various theorems due to
Gromov, Lawson, Grigori Perelman, and John Milnor; see
\cite{kn:arrow} for the details. The answer is ``no": all
non-toral compact three-dimensional manifolds are unsatisfactory
either because they are like the sphere (that is, they support
arbitrarily irregular metrics with non-negative scalar curvature)
or because, no matter how they are deformed, they have \emph{no}
metric of non-negative scalar curvature (so that they do not
permit a universe to be created along them in the first place).
Thus, according to the ideas advanced here, we can say that the
construction of Ooguri et al. \cite{kn:ooguri} leads precisely to
the \emph{only} universes which can have an Arrow.

This, too, is a very remarkable conclusion. For we should not
imagine that a torus with a generic geometry will ``look very
different" from any other space endowed with its generic geometry.
(Think of an extremely distorted two-dimensional torus: the
``handle" distinguishing it from a sphere could be very small and
inconspicuous, and the fact that it has exactly \emph{one} handle
will be far from obvious.) Thus we see that the Arrow is an
\emph{intrinsically global} phenomenon. That is, in order to
determine whether an Arrow will arise in a given set of
circumstances, it is not enough to know what is happening in the
immediate vicinity of an observer; one has to examine the entire
spatial section.

Now in this connection it is important to understand that, while
Eve is spatially finite, her babies are \emph{not}. This strange
fact can be roughly explained as follows. We said that a
Coleman-De Luccia baby universe expands outward into the mother
universe at a rate approaching the speed of light. This means
that, in the manner familiar from special relativity theory, there
is an increasing disparity between lengths in the spatial sections
inside the baby and those of the outside universe. Eventually this
effect becomes so extreme that it becomes possible to fit an
infinite space inside the mother spacetime ---$\,$ the point being
that, as usual in relativity, mother and child have different ways
of slicing spacetime into spatial sections. In fact, ignoring
perturbations, one finds that the babies have spatial sections
which are copies of an infinite space with constant negative
curvature; it is well known that this shape can fit inside the
forward light cone of any point in Minkowski spacetime, and it
fits inside the baby universe in much the same way. (We discussed
the consequences of this shape for the baby's spatial slices in
sections 3.3 and 4.3.) Even if we do \emph{not} ignore
perturbations, this will only change the geometry, \emph{not the
topology}, of the spatial sections of a baby.

The global structure of the babies is therefore quite different to
that of Eve. The topology of the spatial sections near their birth
(and thereafter) is that of ordinary infinite space, which is
certainly not constrained by any result analogous to the SYGLB
theorem. Hence, our argument showing that Eve has an Arrow will
not work for the babies. This agrees with our thermodynamic
argument to the effect that the babies can only have an Arrow if
they can \emph{inherit} one, ultimately from Eve.

The fact that the Arrow is a global phenomenon also solves one of
the most puzzling questions about time: why should the destruction
of spacetime \emph{not} be associated with ``special", low-entropy
conditions?

\subsubsection*{{\textsf{4.5. Against ``Final" Conditions}}}

We agreed in section 2.4 that a successful theory of the Arrow
must explain not just how it arose at the creation, but also what
happens to it over the full span of history. Let us see how this
works in the theory we have developed here.

A spacetime (like Eve) with a positive vacuum energy cannot be
destroyed after it is created, because it never stops expanding.
Eventually, however, Eve will have descendants which do destroy
themselves. This happens, as we discussed in section 4.3, when a
universe in the Landscape ``overshoots" as the value of its vacuum
energy decreases, so that the value becomes \emph{negative}. What
happens to the Arrow, assuming that this universe has inherited
one, near the ``destruction end" of such a universe?

The answer is simple: a universe of this kind will have spatial
sections which, like those of any other baby universe, do
\emph{not} have toral topology. Thus we \emph{cannot} apply the
SYGLB theorem to the spacelike slice along which this universe
ceases to exist. We therefore have no reason to expect that the
geometry of this slice will be in any way ``special".

The only reason we believe in the existence of ``special
conditions" at the creation of our Universe is that Nature gives
us no option: the tendency of entropy to increase is a fundamental
fact, confirmed ``experimentally" at each moment by all of us. By
sharp contrast, there is no evidence whatever for special
conditions at any time in our future. Since we now have a theory
in which an asymmetry between creation and destruction is natural
and inevitable, the parsimonious course is to assume that there
are no special conditions at any time in our future. The fate of
the Universe, from start to finish, is determined by special
\emph{initial} conditions alone. In the case of universes in the
Landscape which overshoot into negative vacuum energy, we can
expect that, as they contract, their anisotropies will grow
rapidly, and that their final spatial geometry will be extremely
distorted, in agreement with the second law of thermodynamics. In
short: the global nature of the Arrow ensures that it points in
the same direction throughout the Landscape, even in universes
which destroy themselves.

The only other regions where spacetime is ``destroyed" is inside
black holes \cite{kn:price}. The local circumstances around black
hole singularities (\cite{kn:waldbook}, Chapter 12) are indeed
somewhat like those near to the singularity in a collapsing
universe. The resemblance does not stop there, however: the
topology of the ``final" spatial section inside a black hole is
\emph{also} non-toral, even if the black hole arises from the
collapse of a star in a toral cosmology. As before, it can be
shown that there is no analogue of the SYGLB theorem for this
topology; so there is no reason to think that the Arrow behaves in
any unusual way inside black holes\footnote{This comment applies
to universes like our own, with positive vacuum energy. The
situation in the negative vacuum energy case is more complex and
remains to be fully understood.}. Again, the global nature of the
Arrow is crucial here.

It is clear from this discussion that the only way ``special
conditions" could ever arise in connection with destruction is if
a \emph{toral} universe is destroyed. We stress that, in our
picture of the Arrow, Eve is the \emph{only} toral universe, and
Eve is never destroyed. In that sense, ``destruction along a
torus" is not an issue. Nevertheless, the logic of our model
demands that it should never, even hypothetically, be possible to
enforce low-entropy conditions at a surface where a universe is
being destroyed; otherwise an observer might find ``the
destruction of the universe" to be in his \emph{past}. That would
suggest that the theory is internally inconsistent.

The proof that this cannot happen is instructive. Let us suppose
that we have a toral universe which is contracting towards what
would be, classically, a final singularity. Now in discussing this
situation, we are eager to avoid the charge of cosmic hypocrisy,
so let us treat it, as far as possible, in exactly the same way as
we treated the creation of a universe. We shall assume, then, that
when this universe contracts to about the string length scale, the
``Casimir-like" negative energy becomes important, violating the
Null Energy Condition, and invalidating the Singularity Theorems.
(Unlike negative vacuum energy, the ``Casimir-like" field has
negative pressure, and so its gravitational field is
\emph{repulsive}; thus it has the power to stop the contraction.)
This is how $\varrho$ can be negative, in accordance with our
mathematical formulation of ``destruction".

But with \emph{negative} $\varrho$ inserted into the second member
of (\ref{eq:L}), we find that our earlier argument simply cannot
get started: $R(h)$ is equal to an expression which is a sum of a
non-negative term with one which is negative. Such a sum could be
of either sign, so we cannot apply the SYGLB theorem here, except
to put bounds on the squared extrinsic curvature (which cannot,
for example, at every point exceed the local value of
$16\pi\,|\varrho|$). Thus we can put very mild constraints on the
rate at which the universe rushes to its doom, but we cannot
constrain its geometry when it \emph{arrives} there.

To see this in detail, let us assume for simplicity that the
extrinsic curvature of the ``final" boundary does vanish. Then the
second member of (\ref{eq:L}) requires the scalar curvature to be
negative everywhere. We know that demanding \emph{non-negative}
scalar curvature imposes strong conditions on the boundary
geometry. What are the consequences of having scalar curvature
which \emph{is} negative? Is there a symmetry, in the space of all
metrics on the torus, between metrics with scalar curvatures of
opposite signs? Here we can use the following theorem, which is an
immediate consequence of the Kazdan-Warner theorem mentioned
earlier \cite{kn:kazdan}:

\bigskip
\noindent \textsf{THEOREM}: Let $M$ be \emph{any} compact manifold
of dimension at least 3, and let $f$ be \emph{any} scalar function
on $M$ such that $f$ is negative somewhere on $M$. Then there
exists a Riemannian metric on $M$ having $f$ as its scalar
curvature.
\bigskip

We see that negative scalar curvature is radically different to
the positive case. Non-negative scalar curvature imposes extremely
strong conditions; negative scalar curvature imposes none at all.
Fix any negative function on the torus, no matter how convoluted,
and there will be some (equally convoluted) geometry such that the
scalar curvature is equal to the given function. Thus there is no
reason whatever to expect the ``destruction" surface to have a
``special" geometry or for any physical field to be in a
``special" state. Entropy will \emph{always} be high when a
universe is destroyed, even in the case of toral topology. Hence
our theory is internally consistent.

What, then, is the ultimate origin of the strange asymmetry
between the the size of the entropy at the creation and
destruction of a universe, even when the topology is toral? The
answer is found partly in the detailed structure of the
general-relativistic initial value constraints, but mainly in the
extreme asymmetry of the space of all metrics on the
three-dimensional torus. In this vast space, the set of metrics
with non-negative scalar curvature is absurdly small compared to
the set of metrics with negative scalar curvature: this is the
content of the SYGLB theorem and of the Kazdan-Warner theorem.
This \emph{geometrically} special condition is turned into the
\emph{physically} special state of the early universe by the
gravitational initial value constraints and by the way they
control the initial state of the inflaton.

\addtocounter{section}{1}
\section*{\large{\textsf{5. Conclusion: The Ageing Of The Multiverse }}}
For each one of us, the passing of time is a basic aspect of
experience: the transience of life is intimately related to its
meaning. In the theory presented here, this phenomenon is a relic
of conditions which were established even before our Universe was
born, a souvenir of the emergence of time itself.

What we have found here may be described as the final triumph of
the second law of thermodynamics. The latter is now seen to rule
not just our Universe, but the entire Landscape; in particular,
the birth of new universes occurs in a manner that is strictly in
accord with the second law; our Universe has low entropy now
because it was preceded by a universe which had still lower
entropy, and so on back to Eve. In short, the multiverse
\emph{ages}.

What remains is to develop a \emph{quantitative} theory of this
ageing process. This will involve (at least) three major steps.

First, we need a precise analysis of the quantum-mechanical process
which breaks the perfect symmetry of the spatial sections of the Eve
universe. Unfortunately, this process is not yet fully understood at
a fundamental level: the perturbations seem to occur on length
scales even shorter than the Planck scale. (See \cite{kn:dyson} and
\cite{kn:brandenberger} for discussions of this ``transplanckian
modes" problem.) More generally, one will want to have a better
understanding of the ``emergence" of time in terms of the
quantum-mechanical version of the main initial-value constraint,
namely the Wheeler-De Witt equation, possibly along the lines
suggested in \cite{kn:hat}. Perhaps a deeper understanding of the
theory of Ooguri et al. \cite{kn:ooguri} will help here.

Secondly, we need a much better understanding of the way the
geometric entropy grows during the process of the birth of a baby
universe, and of the way this fits together with ideas about how
baby universes inflate \cite{kn:frei}.

Finally, we need a clearer picture of the \emph{rate} at which baby
universes nucleate, so that we can predict the conditions in the Eve
universe at the time when this happens. Note that transitions from
larger values of the vacuum energy to smaller values can involve
quite large steps, so the number of steps from Eve to our Universe
need not be large; this important fact \cite{kn:TASI} needs to be
taken into account. In addition, recent developments
\cite{kn:tye}\cite{kn:podolsky}\cite{kn:sash} have suggested that
there may exist one or more mechanisms which can greatly shorten the
time needed for transitions to occur. Thus it may be possible for
babies to be born when Eve is still extremely young. Clearly, much
remains to be done.

The long-range hope is that one will eventually be able to
\emph{compute}, using greatly refined theories of Inflation and
the Landscape, a precise theoretical value for Penrose's number
$P$. That would be clear evidence that we have finally understood
the Arrow in all of its aspects.

\addtocounter{section}{1}
\section*{\large{\textsf{Acknowledgements}}}
The author is deeply grateful to Ruediger Vaas for inviting him to
contribute to this volume, thereby providing him with an
opportunity to polish his ideas very considerably, and to Jens
Niemeyer and the other organizers of and participants in the
Wuerzburg meeting on ``Initial Conditions in Cosmology", where
some of these ideas were presented. Discussions with Dr Soon
Wanmei were also very helpful.


\begin{thebibliography}{18}
\bibitem{kn:schellekens}
A.N. Schellekens, The Landscape ``avant la lettre",
arXiv:physics/0604134
\bibitem{kn:landscape} Leonard Susskind, The
Anthropic Landscape of String Theory, arXiv:hep-th/0302219
\bibitem{kn:TASI}
Raphael Bousso, TASI Lectures on the Cosmological Constant,
arXiv:0708.4231
\bibitem{kn:riess}
A. G. Riess et al., Observational evidence from supernovae for an
accelerating universe and a cosmological constant, Astron. J. 116
(1998) 1009-1038, \x astro-ph/9805201
\bibitem{kn:perlmutter}
S. Perlmutter et al., Measurements of Omega and Lambda from 42
high-redshift supernovae, Astrophys. J. 517 (1999) 565-586, \x
astro-ph/9812133.
\bibitem{kn:tegmark}
Max Tegmark, The Mathematical Universe, arXiv:0704.0646
\bibitem{kn:penrose}
R. Penrose, Singularities and Time-Asymmetry, pp. 581-638 in
\emph{General Relativity: An Einstein Centenary Survey}, eds S W
Hawking, W Israel, Cambridge University Press, Cambridge, 1979
\bibitem{kn:lebowitz}
Joel L. Lebowitz, Time's Arrow and Boltzmann's Entropy, pp.
131-146 in \emph{Physical Origins of Time Asymmetry,} eds J.J.
Halliwell, J. Perez-Mercader, W.H. Zurek, Cambridge University
Press, Cambridge, 1994
\bibitem{kn:zeh}
H. D. Zeh, \emph{The Physical Basis of The Direction of Time},
Fifth Edition, Springer, Heidelberg, 2007
\bibitem{kn:albrecht}
Andreas Albrecht, Cosmic Inflation and the Arrow of Time, pp.
363-401 in \emph{Science and Ultimate Reality: Quantum Theory,
Cosmology and Complexity}, eds J. D. Barrow, P.C.W. Davies, C.L.
Harper, Cambridge University Press, Cambridge, 2004, \x
astro-ph/0210527
\bibitem{kn:price}
Huw Price, Cosmology, Time's Arrow, and That Old Double Standard,
pp. 66-96 in \emph{Time's Arrows Today}, ed S. Savitt, Cambridge
University Press, Cambridge, 1994, \x gr-qc/9310022; The
Thermodynamic Arrow: Puzzles and Pseudo-puzzles, \x
physics/0402040
\bibitem{kn:dyson}
Lisa Dyson, Matthew Kleban, Leonard Susskind, Disturbing
Implications of a Cosmological Constant, JHEP 0210 (2002) 011, \x
hep-th/0208013
\bibitem{kn:carroll}
Sean M. Carroll, Jennifer Chen, Spontaneous Inflation and the
Origin of the Arrow of Time, \x hep-th/0410270
\bibitem{kn:wald}
Robert M. Wald, The Arrow of Time and the Initial Conditions of
the Universe, \x gr-qc/0507094
\bibitem{kn:arrow}
Brett McInnes, Arrow of Time in String Theory, Nucl. Phys. B782
(2007) 1-25, \x hep-th/0611088
\bibitem{kn:baby}
Brett McInnes, Good Babies vs. Bad Babies; or, Inheriting the
Arrow of Time, arXiv:0705.4141(hep-th)
\bibitem{kn:kiefer}
Claus Kiefer, Quantum Cosmology and the Arrow of Time,
Braz.J.Phys. 35 (2005) 296-299, arXiv:gr-qc/0502016
\bibitem{kn:laura}
R. Holman, L.Mersini-Houghton,  Why the Universe Started from a
Low Entropy State, Phys.Rev. D74 (2006) 123510, \x hep-th/0511102
\bibitem{kn:lindereview}
Andrei Linde, Inflationary Cosmology, arXiv:0705.0164 (hep-th)
\bibitem{kn:tegmark2}
Mark P. Hertzberg, Max Tegmark, Shamit Kachru, Jessie Shelton,
Onur Ozcan, Searching for Inflation in Simple String Theory
Models: An Astrophysical Perspective, arXiv:0709.0002 (astro-ph)
\bibitem{kn:hawkingevap}
S. W. Hawking, Breakdown Of Predictability In Gravitational
Collapse, Phys. Rev. D 14, 2460-2473 (1976)
\bibitem{kn:juan}
Juan M. Maldacena, Eternal Black Holes in AdS, JHEP 0304 (2003)
021, \x hep-th/0106112
\bibitem{kn:boltzmann}
L. Boltzmann, On certain questions of the theory of gases, Nature
51 (1895) 413-415
\bibitem{kn:thursday}
http://en.wikipedia.org/wiki/Last\_Thursdayism
\bibitem{kn:KKLT}
Shamit Kachru, Renata Kallosh, Andrei Linde, Sandip P. Trivedi, de
Sitter Vacua in String Theory, Phys.Rev. D68 (2003) 046005, \x
hep-th/0301240
\bibitem{kn:bouff}
Raphael Bousso, Ben Freivogel, A paradox in the global description
of the multiverse, JHEP 0706 (2007) 018 arXiv:hep-th/0610132
\bibitem{kn:krauss}
Irit Maor, Lawrence Krauss, Glenn Starkman, Anthropics and
Myopics: Conditional Probabilities and the Cosmological Constant,
arXiv:0709.0502
\bibitem{kn:davies}
P.C.W. Davies, Stirring Up Trouble, pp. 119-130 in \emph{Physical
Origins of Time Asymmetry,} eds J.J. Halliwell, J. Perez-Mercader,
W.H. Zurek, Cambridge University Press, Cambridge, 1994
\bibitem{kn:ross}
Simon F. Ross, Black hole thermodynamics, arXiv:hep-th/0502195
\bibitem{kn:gibhawk}
G.W. Gibbons, S.W. Hawking, Cosmological Event Horizons,
Thermodynamics, and Particle Creation, Phys.Rev.D15 (1977)
2738-2751
\bibitem{kn:hawking}
S.W. Hawking , The Arrow Of Time In Cosmology, Phys.Rev.D32 (1985)
2489- 2495
\bibitem{kn:laf}
S.W. Hawking, R. Laflamme, G.W. Lyons, The Origin of time
asymmetry, Phys.Rev.D47 (1993) 5342-5356, \x gr-qc/9301017
\bibitem{kn:gibhart}
G.W. Gibbons, J.B. Hartle, Real Tunneling Geometries and the
Large-Scale Topology of the Universe, Phys.Rev. D42 (1990)
2458-2468
\bibitem{kn:maldacena}
Brett McInnes, Quintessential Maldacena-Maoz Cosmologies, JHEP
0404 (2004) 036, \x hep-th/0403104
\bibitem{kn:uggla}
Woei Chet Lim, Henk van Elst, Claes Uggla, John Wainwright,
Asymptotic isotropization in inhomogeneous cosmology, Phys.Rev.
D69 (2004) 103507, arXiv:gr-qc/0306118
\bibitem{kn:gibsol}
G. W. Gibbons, S. N. Solodukhin, The Geometry of Large Causal
Diamonds and the No Hair Property of Asymptotically de-Sitter
Spacetimes, arXiv:0706.0603
\bibitem{kn:loeb}
Kentaro Nagamine, Abraham Loeb, Future Evolution of Nearby
Large-Scale Structure in a Universe Dominated by a Cosmological
Constant, New Astron. 8 (2003) 439-448, arXiv:astro-ph/0204249
\bibitem{kn:fargu}
Edward Farhi, Alan H. Guth, An Obstacle To Creating A Universe In
The Laboratory, Phys.Lett.B183 (1987) 149-155
\bibitem{kn:aguirrejohn}
Anthony Aguirre, Matthew C. Johnson, Two Tunnels to Inflation,
Phys.Rev. D73 (2006) 123529, arXiv:gr-qc/0512034
\bibitem{kn:hubeny}
Ben Freivogel, Veronika E. Hubeny, Alexander Maloney, Robert C.
Myers, Mukund Rangamani, Stephen Shenker, Inflation in AdS/CFT,
JHEP 0603 (2006) 007, \x hep-th/0510046
\bibitem{kn:turok}
Joel K. Erickson, Daniel H. Wesley, Paul J. Steinhardt, Neil
Turok, Kasner and Mixmaster behavior in universes with equation of
state w $\geq$ 1, Phys.Rev. D69 (2004) 063514, \x hep-th/0312009
\bibitem{kn:podolsky2}
D. Podolsky, General asymptotic solutions of the Einstein equations
and phase transitions in quantum gravity, arXiv:0704.0354 (hep-th)
\bibitem{kn:fischler}
W. Fischler, D. Morgan, J. Polchinski, Quantization of
false-vacuum bubbles: A Hamiltonian treatment of gravitational
tunneling, Phys. Rev. D42 (1990) 4042-4055
\bibitem{kn:vacha} Tanmay Vachaspati, On
Constructing Baby Universes and Black Holes, arXiv:0705.2048
\bibitem{kn:page}
Don N. Page, Is Our Universe Decaying at an Astronomical Rate?, \x
hep-th/0612137
\bibitem{kn:deluccia}
Sidney R. Coleman, Frank De Luccia, Gravitational Effects On And
Of Vacuum Decay, Phys.Rev. D21 (1980) 3305-3315
\bibitem{kn:gagv}
Jaume Garriga, Alan H. Guth, Alexander Vilenkin, Eternal
inflation, bubble collisions, and the persistence of memory, \x
hep-th/0612242
\bibitem{kn:aguijohsho}
Anthony Aguirre, Matthew C Johnson, Assaf Shomer, Towards observable
signatures of other bubble universes, Phys.Rev.D76 (2007) 063509,
arXiv:0704.3473
\bibitem{kn:vilenkin}
A. Vilenkin, Creation of Universes from Nothing, Phys.Lett.B117
(1982) 25-28
\bibitem{kn:hartle}
J.B. Hartle and S.W. Hawking, Wave Function of the Universe,
Phys.Rev.D28 (1983) 2960-2975
\bibitem{kn:polchinski}
Joseph Polchinski, The Cosmological Constant and the String
Landscape, \x hep-th/0603249
\bibitem{kn:vaas}
Ruediger Vaas, Time before Time - Classifications of universes in
contemporary cosmology, and how to avoid the antinomy of the
beginning and eternity of the world, Bild.Wiss. 10 (2004) 32-41,
arXiv:physics/0408111
\bibitem{kn:mars}
Marc Mars, José M. M. Senovilla, Raül Vera, Is the accelerated
expansion evidence of a forthcoming change of signature?,
arXiv:0710.0820 (gr-qc)
\bibitem{kn:ooguri}
Hirosi Ooguri, Cumrun Vafa, Erik Verlinde, Hartle-Hawking
Wave-Function for Flux Compactifications: The Entropic Principle,
Lett.Math.Phys. 74 (2005) 311-342 , \x hep-th/0502211
\bibitem{kn:OVV}
Brett McInnes, The Geometry of The Entropic Principle and the
Shape of the Universe, JHEP 10 (2006) 029, \x hep-th/0604150
\bibitem{kn:polchbook}
J. Polchinski, {\em String Theory}, Cambridge University Press,
Cambridge, 1998
\bibitem{kn:brandenberger}
Robert H. Brandenberger, Conceptual Problems of Inflationary
Cosmology and a New Approach to Cosmological Structure Formation,
Lect.Notes Phys.738 (2008) 393, arXiv:hep-th/0701111
\bibitem{kn:lindetypical}
Andrei Linde, Creation of a Compact Topologically Nontrivial
Inflationary Universe, JCAP 0410 (2004) 004, \x hep-th/0408164
\bibitem{kn:kobayashi}
S. Kobayashi, K. Nomizu {\em Foundations of Differential Geometry
II}, Interscience, New York, 1969
\bibitem{kn:waldbook}
Robert M. Wald, \emph{General Relativity}, Chicago University
Press, Chicago, 1984
\bibitem{kn:kazdan}
Kazdan, Jerry L., Warner, F. W., Existence and conformal
deformation of metrics with prescribed Gaussian and scalar
curvatures. Ann. of Math. 101 (1975) 317-331
\bibitem{kn:lawson}
H. Blaine Lawson and Marie-Louise Michelsohn, \emph{Spin
Geometry}, Princeton University Press, Princeton, 1990
\bibitem{kn:coule}
D.H. Coule, Quantum Cosmological Models, Class.Quant.Grav. 22
(2005) R125-2308, \x gr-qc/0412026
\bibitem{kn:nimah1}
Nima Arkani-Hamed, Sergei Dubovsky, Alberto Nicolis, Giovanni
Villadoro, Quantum Horizons of the Standard Model Landscape, JHEP
06 (2007) 078, arXiv:hep-th/0703067
\bibitem{kn:nimah2}
Nima Arkani-Hamed, Sergei Dubovsky, Alberto Nicolis, Enrico
Trincherini, Giovanni Villadoro, A Measure of de Sitter Entropy
and Eternal Inflation, JHEP 0705 (2007) 055, arXiv:0704.1814
(hep-th)
\bibitem{kn:hat}
Leonard Susskind, The Census Taker's Hat, arXiv:0710.1129 (hep-th)
\bibitem{kn:frei}
Ben Freivogel, Matthew Kleban, Maria Rodriguez Martinez, Leonard
Susskind, Observational Consequences of a Landscape, JHEP 0603
(2006) 039, \x hep-th/0505232
\bibitem{kn:tye}
S.-H. Henry Tye, A New View of the Cosmic Landscape,
arXiv:hep-th/0611148
\bibitem{kn:podolsky}
D. Podolsky, K. Enqvist, Eternal inflation and localization on the
landscape, arXiv:0704.0144 (hep-th)
\bibitem{kn:sash}
Sash Sarangi, Gary Shiu, Benjamin Shlaer, Rapid Tunneling and
Percolation in the Landscape, arXiv:0708.4375 (hep-th)




\end{thebibliography}
\end{document}